\documentclass[twocolumn]{aa}
\usepackage{psfig}
\usepackage{graphicx}
\begin{document}
%
%%%%%%%%%%%%%%%%%%%%%%%%%%%%%%%%%%%%%%%%%%%%%%%%%%%%%%%%%%%%%%%%%%
%
%\thesaurus{11(11.06.1;  % Galaxies: formation,
%              11.06.2;  % Galaxies: fundamental parameters,
%              11.16.1;  % Galaxies: photometry,                    
%              11.19.2;  % Galaxies: spiral,
%              11.19.7;  % Galaxies: statistics,
%              11.19.6)} % Galaxies: structure.
%%%%%%%%%%%%%%%%%%%%%%%%%%%%%%%%%%%%%%%%%%%%%%%%%%%%%%%%%%%%%%%%%%
%
   \title{Modelling the spectral energy distribution of galaxies. IV}
   \subtitle{Correcting apparent disk scalelengths and central surface
   brightnesses for the effect of dust at optical and near-infrared 
   wavelengths.}

   \author{C. M\"ollenhoff
          \inst{1}
          \and
          C. C. Popescu\inst{2,3,4}
          \and
          R. J. Tuffs\inst{2,3}
          }

   \offprints{C.C. Popescu \& R.J. Tuffs}

   \institute{Zentrum f\"ur Astronomie der Universit\"at Heidelberg, 
Landessternwarte, K\"onigstuhl 12, D-69117 Heidelberg, Germany\\
      \email{cmoellen@lsw.uni-heidelberg.de}
   \and
       Max-Planck-Institut f\"ur Kernphysik, Saupfercheckweg 1,
        D-69117 Heidelberg, Germany\\
             \email{Cristina.Popescu@mpi-hd.mpg.de\,\,\,Richard.Tuffs@mpi-hd.mpg.de}
\and
The Observatories of the Carnegie Institution of Washington, 813 Santa Barbara
St., Pasadena, CA 91101, USA
  \and 
     Research Associate, The Astronomical Institute of the Romanian
           Academy, Str. Cu\c titul de Argint 5, Bucharest, Romania     
             }

   \date{Received 20/12/2005; accepted 16/06/2006}

\abstract{We present corrections for the change in the apparent scalelengths,
central surface brightnesses and axis ratios due to the presence of dust in 
pure disk galaxies, as a function of inclination, central face-on opacity in
the B-band (${\tau}^{f}_{\rm B}$) and 
wavelength. The correction factors were derived from simulated images of disk
galaxies created using geometries for stars and dust which can reproduce the
entire spectral energy distribution from the ultraviolet (UV) to the
Far-infrared (FIR)/submillimeter (submm) and can also account for the 
observed surface-brightness distributions in both the optical/Near-infrared 
and FIR/submm. We found that dust can significantly affect both the scalelength
and central surface brightness, inducing variations in the apparent to 
intrinsic
quantities of up to 50$\%$ in scalelength and up to 1.5 magnitudes in 
central surface brightness. We also identified some astrophysical effects for 
which, although the absolute effect of dust is non-negligible,
the predicted variation over a likely range in opacity is relatively small, 
such that
an exact knowledge of opacity is not needed. Thus, for a 
galaxy 
at a typical inclination of $37^{\circ}$ and having any 
${\tau}^{f}_{\rm B}>2$, the 
effect of dust is to increase the scalelength in B relative to that in I by a 
factor of $1.12 \pm 0.02$  and to change the B-I central colour by 
$0.36\pm 0.05$ magnitudes. Finally we use the model to analyse the observed 
scalelength ratios between B and I for a sample of disk-dominated spiral 
galaxies, finding that the tendency for apparent scalelength to increase with
decreasing wavelength is primarily due to the effects of dust.

}

\maketitle

\titlerunning{Correcting apparent disk scalelengths and central surface
  brightnesses for the effect of dust.} 
\authorrunning{C. M\"ollenhoff, C. Popescu, R. Tuffs}
%
%%%%%%%%%%%%%%%%%%%%%%%%%%%%%%%%%%%%%%%%%%%%%%%%%%%%%%%%%%%%%%%%%%%
%%%%%%%%%%%%%%%%%%%%%%%%%%%%%%%%%%%%%%%%%%%%%%%%%%%%%%%%%%%%%%%%%%%
\section{Introduction}
\label{intro}

A primary goal of modern studies of star-forming galaxies is to 
understand how these systems were assembled over cosmic time. If the disks of 
spiral galaxies grow from the inside out, as predicted by semi-analytical
hierarchical models for galaxy formation (e.g. Mo, Mao \& White 1998),
one would predict the stellar populations to be younger and have lower
  metallicity in the outer disk than
 in the inner disk, such that local universe galaxies should be intrinsically 
larger at the shorter wavelengths where light from the young stellar 
population is more prominent. For the 
same reason one would expect the intrinsic sizes of spiral disks to be larger 
at the current epoch than at higher redshift. Observationally, such predictions
can be tested in two ways. One way is to compare the spatial distribution of 
the constituent stellar populations at different wavelengths, for local 
universe galaxies. Another way is to look for structural differences in 
galaxies observed at different cosmological epochs, at the same rest frame 
wavelength. Both methods require an analysis of the surface-brightness 
distribution of spiral galaxies in the optical and near-infrared (NIR) to
quantify the distribution of starlight, for example by deriving the 
scalelength of the disk. This is done by fitting observed 
images with models for the surface-brightness distribution of stellar light, 
whereby the disk component is usually specified by an exponential distribution.
 The derived exponential scalelengths can either 
be intercompared between different wavelengths for local universe galaxies
(Elmegreen \& Elmegreen 1984, Peletier et al. 1994, Evans 1994, 
De Jong 1996a,b, de Grijs 1998, Cunow 1998, 2001, 2004, 
MacArthur et al. 2003, M\"ollenhoff 2004), 
or between galaxies at different redshifts at a given wavelength (Lilly et
al. 1998, Simard et al. 1999, Ravindranath et al. 2004, Trujillo \& Aguerri
2004, Trujillo et al. 2005, Barden et al. 2005). However, 
the appearance of disk galaxies is strongly affected by dust and this effect 
is different at different wavelengths and for different opacities.
This has consequences not only for the derivation of the variation of 
intrinsic scalelength with wavelength, but also for the variation of
intrinsic scalelength with cosmological epoch, since the opacity of disk
galaxies is expected to have been systematically higher in the past (
  e.g. Dwek 1998, Pei et al. 1999).

The effect of dust on the observed scalelengths and central
surface brightnesses of disk galaxies has been previously modelled by 
Byun et al. (1994), Evans (1994) and Cunow (2001). By means of radiative
transfer calculations these works investigated the dependence of this effect
on star/dust geometry, opacity, inclination and wavelength. Recently, a better
knowledge of the star/dust geometry has been obtained through a joint
consideration of the direct starlight, emitted in the ultraviolet
(UV)/optical/NIR, and of the starlight which is re-radiated 
in the Far-infrared (FIR)/submillimeter (submm).
In a series of papers devoted
to modelling the spectral energy distributions (SEDs) we derived geometries 
of the distribution of stellar light and dust that are 
successful in reproducing not only the observed integrated SEDs, but also the 
observed radial profiles both in the optical/NIR 
and 
FIR/submm (Popescu et al. 2000; hereafter Paper~I, Misiriotis et al. 2001;
hereafter Paper~II, Popescu et al. 2004, see also Popescu \& Tuffs 2005 
\footnote{A simplified version of this geometrical prescription
has been applied by Misiriotis et al. (2004) to fit the FIR SEDs of 
bright IRAS galaxies.}). 

In this paper, the fourth in this series, we use 
these derived distributions of stars and dust to obtain a new
quantitative measure of the effect of dust on the observed photometric
parameters in the optical wavebands. Furthermore, because it is no longer 
necessary to explore a wide
 range of star/dust geometries, we are also able to
systematically explore the full parameter space in opacity, inclination and
wavelength, and tabulate the results in a form convenient for the use of the 
community. 
We give quantitative measures of the change in the 
apparent scalelength, central surface brightnesses and inclination of disk 
galaxies due to the presence of dust.\footnote{ 
These corrections are only valid
  for normal disk galaxies and are not applicable to systems with different
  star/dust geometries such as starburst or dwarf galaxies.} All these changes 
are expressed as 
the ratio of the apparent quantity (i.e. that obtained by fitting images of 
dusty disks with pure exponential disks) to the intrinsic quantity (i.e. that 
which would be obtained in the absence of dust). These corrections have been 
derived from a subset of the simulated images (those in the optical bands) 
presented in Tuffs et al. (2004; hereafter Paper~III). 

In Sect.~2 we give a brief description of the distributions of stars and dust
used in the simulated images. In Sect.~3 we specify the fitting procedure used
to extract apparent scalelengths, central surface brightnesses and axis ratios
from the simulated images. These quantities are tabulated in Sect. 4,
where we also describe and explain their dependence on opacity, inclination 
and wavelength due to the effect of dust. In Sect.~5 we examine the impact
of these new results on our ability to derive quantities of astrophysical
interest from optical observations and give in Sect.~6 a specific example of the
determination of the variation of intrinsic scalelength with wavelength for 
local universe galaxies. A summary of the paper is given in Sect.~7. 
%%%%%%%%%%%%%%%%%%%%%%%%%%%%%%%%%%%%%%%%%%%%%%%%%%%%%%%%%%%%%%%%%%%

%%%%%%%%%%%%%%%%%%%%%%%%%%%%%%%%%%%%%%%%%%%%%%%%%%%%%%%%%%%%%%%%%%%%
%%%%%%%%%%%%%%%%%%%%%%%%%%%%%%%%%%%%%%%%%%%%%%%%%%%%%%%%%%%%%%%%%%%%
\section{The simulated images}
\label{simulation}

In Paper~III we presented simulated images for the diffuse component of a
spiral galaxy. This diffuse component is comprised of a diffuse old stellar 
population and associated dust and a
diffuse young stellar population and associated dust. 
Direct evidence for the existence of the diffuse dust disk is provided by
 images of FIR emission in spiral galaxies (for example in M~31 by 
Haas et al. 1998 and in M~33 by Hippelein et al. 2003), which clearly show a
diffuse disk of cold dust emission prominent at 170\,${\mu}$m. 
FIR measurements of a statistically unbiased optically selected sample 
of gas-rich galaxies (Tuffs et al. 2002a,b) have shown that this
diffuse cold dust emission component is ubiquitous along
the Hubble sequence of late-type galaxies, and carries the bulk of
the dust luminosity for most of these systems (Popescu et al. 2002). It is thus
responsible for most of the attenuation of the stellar light and the associated
modification of disk brightness and scalelength measured in the optical, which
is the subject of this paper.

We know that in reality the apparently diffuse
  dust component may itself have some structure, and that some fraction 
 (currently
  unconstrained by direct observation) may be contained in clumps which are
  externally heated and have no embedded sources, the so-called ``passive
  clumps'' (Popescu \& Tuffs 2005). However,
  providing these passive clumps are optically thin, the attenuation 
  characteristics
  will be almost identical to those of a completely homogeneous distribution
  (Kuchinski et al. 1998, Pierini et al. 2004). If the passive clumps are 
 considered to be optically thick, then the overall attenuation of a galaxy 
can be reduced by up 
to $40\%$ (Bianchi et al. 2000a, Misiriotis \& Bianchi 2002). But, once UV 
photons can no longer penetrate a clump, the photoelectric heating of the gas 
is lost and there is nothing to prevent the clump collapsing to form
stars. This means that in practice optically thick passive clumps should not
exist as stable long-lived structures, but rather as transient precursors of
star-forming clouds. Once the star-formation is underway, the clumps will start
to contain embedded sources, becoming ``active'' clumps 
(Popescu \& Tuffs 2005). In terms of the global attenuation of stellar light in
galaxies, the main effect of active clumps is to locally absorb a fraction of 
the UV output from the embedded massive stars and transform it into warm dust
emission. This is taken into account by the model of Popescu et al. (2000) used
in this series of papers. Active clumps have such a small filling
factor in normal spiral galaxies that their effect on the propagation of the
long-range optical/NIR photons in galaxy disks is negligible. Since in this
paper we only consider the optical/NIR spectral range, it is only necessary
  here to take into account the diffuse dust distribution.

The diffuse old stellar
population has both a ``bulge'' and a ``disk'' component, whereas the
diffuse young stellar population resides only in a ``thin disk''.
% Of the three diffuse geometrical components presented in Paper~III, namely the 
%``disk'', ``the thin disk'' and ``the bulge'', 
Here we only use the disk component, 
since in this paper we are only concerned with the appearance of the disk in the 
optical/NIR range.
The simulations for the disk were performed taking the geometry of 
the diffuse dust to be the superposition of the dust in the disk and the dust
in the thin disk. In other words the attenuated images represent the appearance
of the stellar populations in the disk as seen through the dust in the disk 
and thin disk. 
The stellar emissivity of the disk $\eta$ and the extinction coefficients of 
the dust associated with the disk, $\kappa^{\rm disk}_{\rm ext}$, and with the 
thin disk, $\kappa^{\rm tdisk}_{ext}$, are described by exponentials:

\begin{eqnarray}
\eta(\lambda,R,z) = {\eta^{\rm disk}}(\lambda,0,0) 
\exp \left( - \frac{R}{{h^{\rm disk}_{\rm s}}} - 
\frac{|z|}{{z^{\rm disk}_{\rm s}}} \right)
\end{eqnarray}
\begin{eqnarray}
\kappa^{\rm disk}_{\rm ext}(\lambda,R,z) = {\kappa^{\rm disk}_{ext}
(\lambda,0,0)}\,\exp \left( - \frac{R}{h^{\rm disk}_{\rm d}}- 
\frac{|z|}{z^{\rm disk}_{\rm d}} \right)
\end{eqnarray}
\begin{eqnarray}
\kappa^{\rm tdisk}_{ext}(\lambda,R,z) = 
\kappa^{\rm tdisk}_{ext}(\lambda,0,0)\,\exp 
\left( - \frac{R}{h^{\rm tdisk}_{\rm d}}- \frac{|z|}{z^{\rm tdisk}_{\rm d}} 
\right)
\end{eqnarray}
where $R$ and $z$ are the cylindrical coordinates, 
$\eta^{\rm disk}(\lambda,0,0)$ is the stellar emissivity at the centre of the 
disk, $h^{\rm disk}_{\rm s}$ and $z^{\rm disk}_{\rm s}$ are the scalelength 
and scaleheight of the stellar emissivity of the disk, 
$\kappa^{\rm disk}_{\rm ext}(\lambda,0,0)$ is the extinction coefficient
 at the centre of the disk, $h^{\rm disk}_{\rm d}$ and 
$z^{\rm disk}_{\rm d}$ are the scalelength and scaleheight of the dust 
associated with the disk, $\kappa^{\rm tdisk}_{ext}(\lambda,0,0)$ is the 
extinction coefficient at
the centre of the thin disk and $h^{\rm tdisk}_{d}$ and $z^{\rm tdisk}_{d}$ are
the scalelength and scaleheight of the dust associated with the thin
disk. The values of these parameters are given in Tables~1
and 2 from Paper~III. For the old stellar population and associated dust the
values of these parameters were taken from the modelling of Xilouris
et al. (1999). It should be noted that the thin disk of dust is an
  approximation to the distribution of dust associated with the spiral arms, as
  explained in Paper I, II and III. This approximation will be a better
  representation of spirals having multiple spiral arms covering a large
  fraction of the disk than of spirals having a two arm pattern with low
  coverage of the disk or of spirals with a ring structure. This may introduce
  some noise in the attenuation characteristics, in particular at the longest
  NIR wavelength, where the true relative geometry of stars and dust may differ
  the most from that assumed by the model. However it is at these wavelengths
  that attenuation is the least severe. It should also be noted that the geometry used by our model is 
only valid for massive (non-dwarf) spiral galaxies, for which Dalcanton et al. (2004) have
suggested that disk instabilities lead to a thin
dust layer through collapse and fragmentation of high density gas within spiral
arms.  For low-mass galaxies with circular 
velocities of less than 120\,km/s Dalcanton et al. (2004) have shown that the 
disk has a different structure, with the dust having a larger scaleheight 
with respect to the stars than in more massive, spiral, galaxies.

The simulated images of the disk were calculated 
using the radiative transfer code of Kylafis \& Bahcall (1987), which includes 
anisotropic multiple scattering in which the higher orders of scattered light
are calculated using the method of ``scattered intensities'' (see also 
Kylafis \& Xilouris 2005). The images have a pixel size (equal to the 
resolution) of 0.0066 of the B-band scalelength $h^{\rm disk}_{\rm s}$ and 
were sampled every 5 and 10 pixels in the inner and outer disk,
respectively. They  
extend out to a radius of 4.63 B-band scalelengths $h^{\rm disk}_{\rm s}$,
which is equivalent to 3.31 dust scalelengths $h^{\rm disk}_{\rm d}$. Examples
of simulated disk images are shown in Fig.~1 from Paper~III.

Here we make use of a subset of 280 of the simulated disk images presented in 
Paper~III, spanning 7 values of total central face-on optical depth in B band 
$\tau_{\rm B}^{\rm f}$, 8 inclinations ${i}$ and 5 wavelengths.
For the sampling in 
$\tau_{\rm B}^{\rm f}$ we chose the set of values 
${0.1,0.3,0.5,1.0,2.0,4.0,8.0}$, which range from extremely optically thin to
moderately optically thick cases.
For the sampling in inclination we used $\Delta \cos(i) = 0.1$, and $0 \leq
(1 - \cos(i)) \leq 0.7$, which correspond to 
$i=0, 26, 37, 46, 53, 60, 66, 73^{\circ}$. The limit of $73^{\circ}$ in $i$ was imposed since standard methods to 
derive apparent exponential scalelength from observed 
images are no longer applicable for higher inclinations.
The simulations were performed in the standard optical/NIR bands B, V, I, J,
and K. 
We also used the corresponding intrinsic images of the stellar emissivity 
(as would be observed in the absence of dust).

%%%%%%%%%%%%%%%%%%%%%%%%%%%%%%%%%%%%%%%%%%%%%%%%%%%%%%%%%%%%%%%%%%%%
%%%%%%%%%%%%%%%%%%%%%%%%%%%%%%%%%%%%%%%%%%%%%%%%%%%%%%%%%%%%%%%%%%%%
\section{Derivation of apparent disk scalelengths and central surface
  brightnesses}
\label{diskfit}

\subsection{Formalism}

When extracting apparent scalelengths, central surface brightnesses and 
axis ratios from observed (dust attenuated) images of galaxy disks, the common 
procedure is to fit the observed surface-brightness distribution with the 
brightness distribution corresponding to the projection of an inclined thin 
axisymmetric exponential disk. This brightness distribution takes the form:

\begin{eqnarray}
I_{\rm app} = I_{\rm app}^{\rm c}\,exp(-\frac{\sqrt{(y^2/Q_{app}^2) +
  x^2)}}{R_{\rm app}}) 
\end{eqnarray}
where $I_{\rm app}^{\rm c}$ is the apparent central surface brightness, 
$R_{\rm app}$ is the apparent exponential scalelength, and $Q_{app}$ is the
apparent axis ratio $b/a$ of the inclined disk (equivalent to the
cosine of the apparent inclination of the disk). The coordinate system is
orientated such that galaxies have their major axis along the x-axis.

The strategy we adopt in this paper is to use the same fitting procedure as
used to fit real observations to derive $I_{\rm app}^{\rm c}$, $R_{\rm app}$ and
$Q_{app}$ from our simulated dust attenuated disk images. These quantities 
can then be compared with the corresponding intrinsic quantities 
$I_{\rm 0}^{\rm c}$, $R_{\rm 0}$, and $Q_{0}$, 
where $I_{\rm 0}^{\rm c}$ is the
intrinsic face-on surface brightness, $R_{\rm 0}$ is the intrinsic  exponential
scalelength, and $Q_{0}$ is the intrinsic inclination of the disk. 
The intrinsic quantities are known as input to our simulations or can be 
derived from the simulated dustless face-on images:
$R_{\rm 0}$ is equal to the value of the scalelength of the stellar
emissivity $h_{\rm s}^{\rm disk}$, $Q_{0}$ is $cos(i)$ for
a simulation made with inclination $i$, and $I_{\rm 0}^{\rm c}$ is the
brightness of the central pixel of the simulated dustless face-on image.
The ratios $I_{\rm app}^{\rm c}$/$I_{\rm 0}^{\rm c}$, $R_{\rm app}$/$R_{\rm
  0}$, and $Q_{app}$/$Q_{0}$ can then be used to quantify the discrepancy
between intrinsic and perceived quantities arising from the presence of dust.

%================================================================

%================================================================

%%%%%%%%%%%%%%%%%%%%%%%%%%%%%%%%%%%%%%%%%%%%%%%%%%%%%%%%%%%%%%%%%%%
\subsection{Technical procedure of the fit}
\label{procedure}

The simulated disk images containing 700 x 1400 pixels and covering half a
galaxy were first converted into full images of galaxies containing 1399x1400
pixels. The sky-value was zero by construction. The maximum of the light 
distribution was determined and was used as the centre of the disk fit.

The two-dimensional surface-brightness function from eq. 4  
was fitted to the brightness distribution of the simulated images following
M\"ollenhoff \& Heidt (2001). 
Fitting $I_{app}$ 
to the two-dimensional surface-brightness distribution results in a 
nonlinear system of equations for the 3 free parameters  $I_{\rm app}^{\rm c}$,
$R_{\rm app}$ and $Q_{app}$.
A Levenberg-Marquardt algorithm was used for the solution of this system
(e.g. Bevington \& Robinson, 1992). 

Several technical steps were necessary as prerequisites for the fit:
The centre of the galaxy was determined in each image and the 
function from eq. 4 was fitted to the whole image. 
Since there were 280 images to fit, a systematic procedure
was necessary. We started in K with the face-on image and the lowest optical
depth ($i = 0^{\circ}$, $\tau_{\rm B}^{\rm f}=0.1$).
A rough guess for the start values of the parameters in the initial 
Levenberg-Marquardt fit was sufficient. The resulting parameters $I_{\rm app}^{\rm c}$,  
$R_{\rm app}$ and $Q_{app}$ were used as start values for the next fits 
($i = 0^{\circ}$, $\tau_{\rm B}^{\rm f} = 0.3, 0.5, 1.0, 2.0, 4.0, 8.0$).
Then the same procedure was repeated with the inclinations $i = 26^{\circ}$ until 
$i = 73^{\circ}$.
Afterwards we proceeded in the same manner for the other filters J, I, V, B.

Fig.~1 shows two examples of simulated disks and of 
residua after subtraction of our photometric fits to these simulated disks. 
For face-on galaxies the fit is perfect and the residuum is 
practically zero. For inclined galaxies  
the stronger forward scattering induces a relative brightening of the near side of the 
disk, leading to an asymmetry around the horizontal axis. Such an asymmetry 
is not considered by our fitting function, therefore the fit to the inclined
galaxies will be imperfect.

To show the details of the photometric fits and of the residua between the
simulated disks and the fits, we plot in  Fig.~2 cuts parallel 
and perpendicular to the major axis. This is shown for simulations done in the
B band, for ${\tau}^{f}_{\rm B}=1.0$ and for three different inclinations, 
$i = 0,46,73^{\circ}$.
Due to the central opacity the disk models are somewhat rounder in the
centre than the corresponding exponential fits. This effect is reflected 
in the central cusp of the residuum (solid bottom line) and increases with 
increasing inclination. Nevertheless the exponential fit is a good
representation of the simulated brightness distribution. The above mentioned 
asymmetry 
due to dust scattering is especially prominent at high inclination in 
the cuts perpendicular to the major axis (bottom right panel of Fig.~2).

%=========== FIGURE: Model  and residuum =============
\begin{figure}
\vbox{
\hbox{\hspace{0cm}              % image row 1
  \psfig{figure=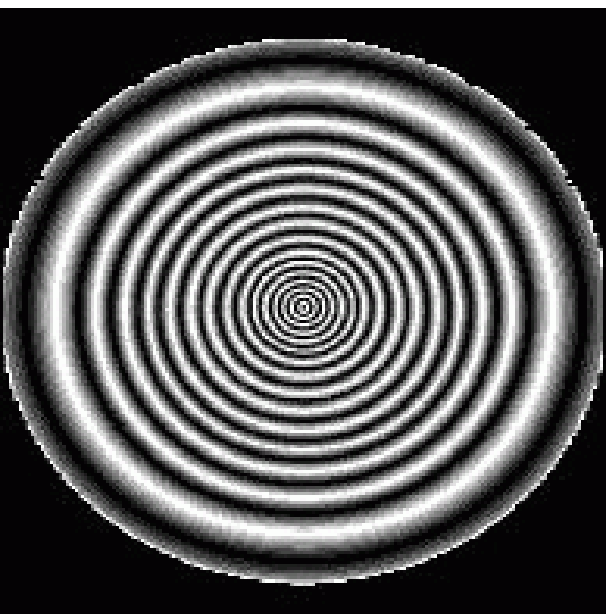,width=4.2cm,clip=}
  \hspace{0.2cm}
  \psfig{figure=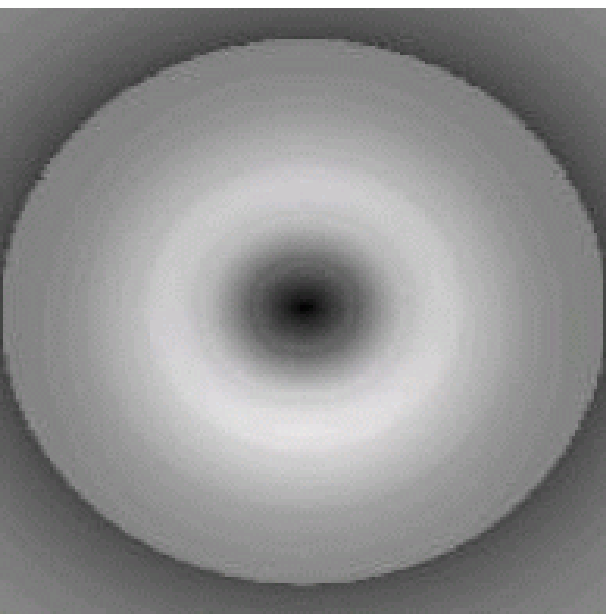,width=4.2cm,clip=}
}
\vspace{0.2cm}
\hbox{\hspace{0cm}              % image row 2
  \psfig{figure=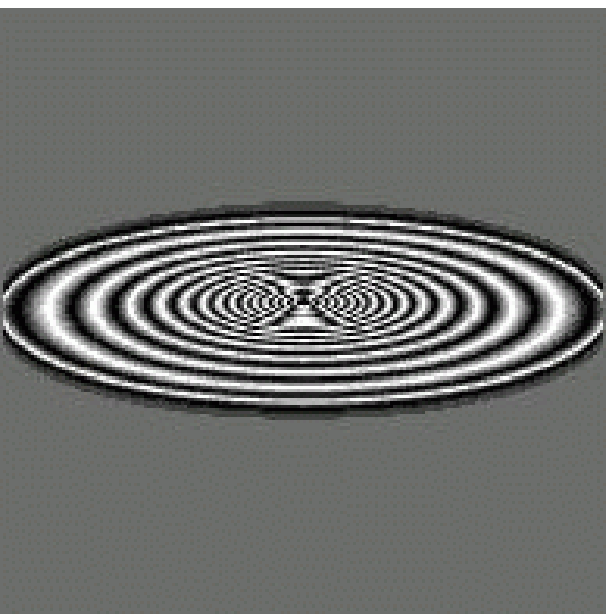,width=4.2cm,clip=}
  \hspace{0.2cm}
  \psfig{figure=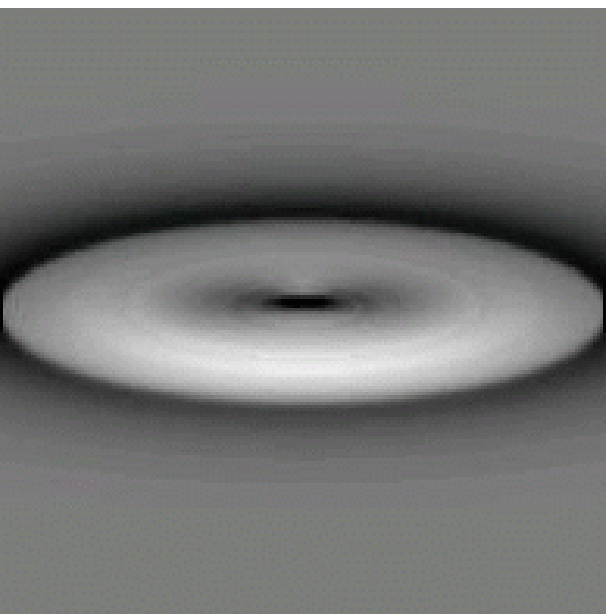,width=4.2cm,clip=}
}}
\caption[]{{\bf Top:} B image of the dusty disk model 
$i = 26^{\circ}$, ${\tau}^{f}_{\rm B} = 8.0$ (left) and residuum after 
subtraction of the photometric fit (right).
{\bf Bottom:} B image of the dusty disk model
$i = 73^{\circ}$, ${\tau}^{f}_{\rm B} = 8.0$ (left) and residuum 
after subtraction of the photometric fit (right).
For a better visibility an alternating intensity table
for the disk models was chosen which produced the isophote-like images.}
\label{figimage} %fig1
\end{figure}

%=========== FIGURE: cuts  =============
\begin{figure}
  \psfig{figure=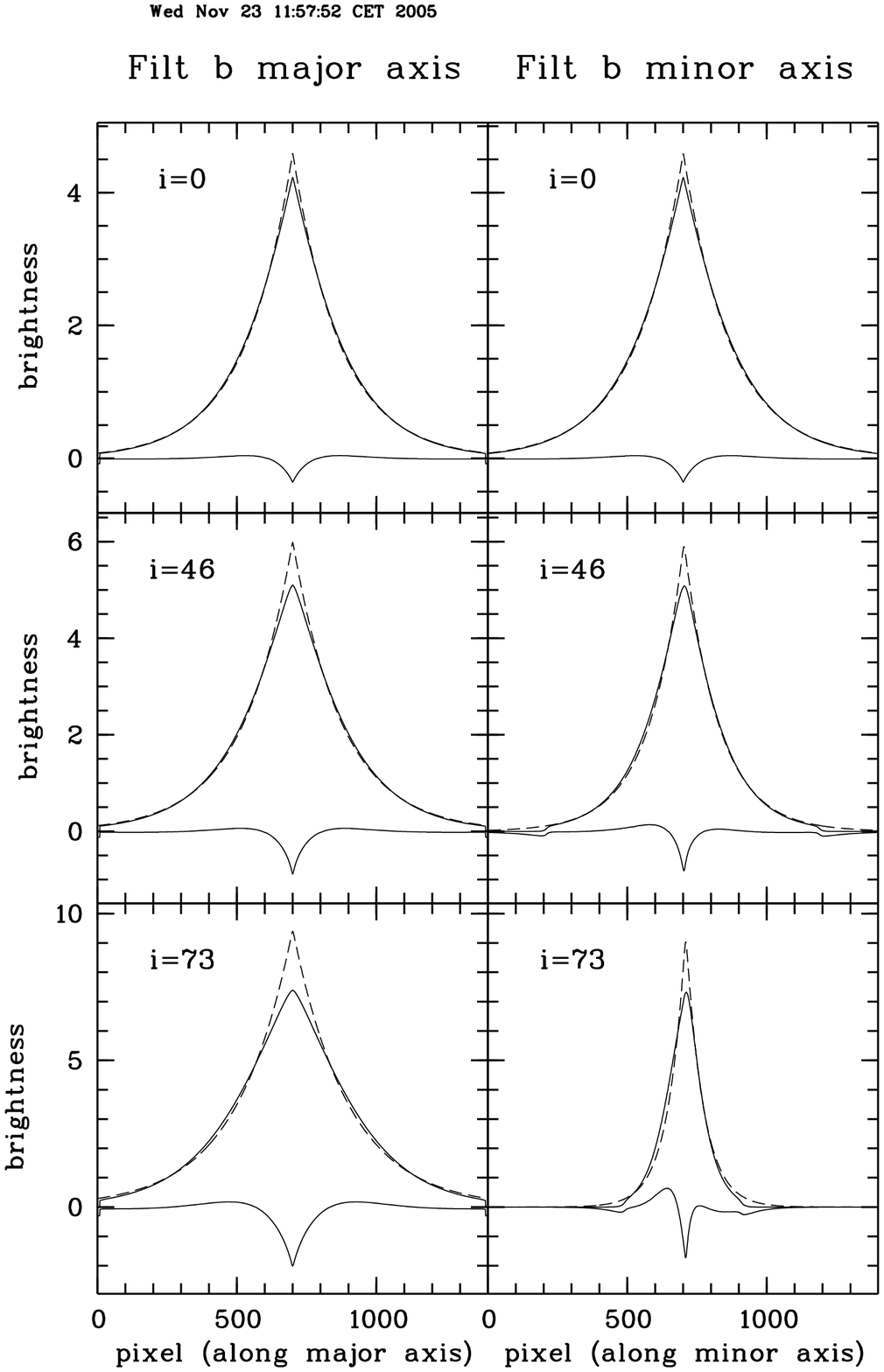,width=8.5cm,clip=}
\caption[]{Behaviour of two-dimensional fits and residua for simulations done
  in the B band, for ${\tau}^{f}_{\rm B}=1.0$,
and for 3 different inclinations: $i = 0$ (top), $i = 46$ (centre),
and $i = 73^{\circ}$ (bottom). Each panel shows
cuts parallel and perpendicular to the major axis, taken through the centres 
of the simulated galaxy (solid lines), of the fit (dashed lines),
and of the residuum (solid lines at bottom).} 
\label{fig6cuts}
\end{figure}

%%%%%%%%%%%%%%%%%%%%%%%%%%%%%%%%%%%%%%%%%%%%%%%%%%%%%%%%%%%%%%%%%%%%
%%%%%%%%%%%%%%%%%%%%%%%%%%%%%%%%%%%%%%%%%%%%%%%%%%%%%%%%%%%%%%%%%%%%
%\section{Results: plots and tables and description}
\section{Results of the photometric fits}
\label{results}

Tables 1 to 5 show the results of the fitting procedure for the
B, V, I, J, and K bands. For each table the first two columns contain 
the input parameters inclination $i$ and central face-on central optical 
depth in the B band ${\tau}^{f}_{\rm B}$ and the last 
four columns
contain the output parameters $Q_{app}$, $R_{\rm app}$/$R_{\rm 0}$, $I_{\rm
  app}^{\rm c}$/$I_{\rm 0}^{\rm c}$ (as defined in Sect.~3.1) and $\Delta SB$.
$\Delta\rm{SB} = -2.5 log(F/F_0)$ is the ratio of the apparent average 
central surface brightness  $F$ to the intrinsic average 
central face-on surface brightness  $F_0$, expressed in magnitudes.
$F$
is an average taken over an elliptical aperture \footnote{We give the average surface brightness in an aperture (in addition
to the surface brightness at the centre position of the fitted template)
because in practice the brightness is blurred over a point spread function.
Therefore it is useful to include a reference aperture which is larger than
typical point spread functions, but smaller than the scalelength of the disk.}
centred on the position of peak
brightness in the simulated dusty image, with a semi-major axis of 
$R_{\rm app}/10$ and an axis ratio of $\cos(i)$.
$F_0$ is an average taken over a circular aperture 
centred on the position of peak
brightness of the face-on dustless simulated image, 
with a radius $R_{\rm 0}/10$.

Here we should mention that the results given here are, as described in
  Sect. 3.2, obtained by fixing the coordinates of the centre of the fit to the
  position of maximum observed brightness. Other fitting procedures
  such as GIM2D (Simard 1998, Marleau \& Simard 1999) or GALFIT (Peng et
  al. 2002)  constrain
  the coordinates of the centre from the moments of the two-dimensional
  intensity distribution, which result in a small shift along the minor axis 
  of the fitted centre
  for opaque high-inclination disks. The effect of this shift on the derivation
  of the central brightness is at most $3\%$. The effect on the derivation of
  the scalelength is negligible, since it depends only to second
  order on the shift.  

In the following subsections we describe the general behaviour of the
structural parameters listed in Tables~1-5 as a function of 
${\tau}^{f}_{\rm B}$, $i$ and wavelength.

%%%%%%%%%%%%%%%%%%%%%%%%%%%%%%%%%%%%%%%%%%%%%%%%%%%%%%%%%%%%%%%%%%%
%=========== FIGURE: Rdn over tau  =============
\begin{figure}
  \psfig{figure=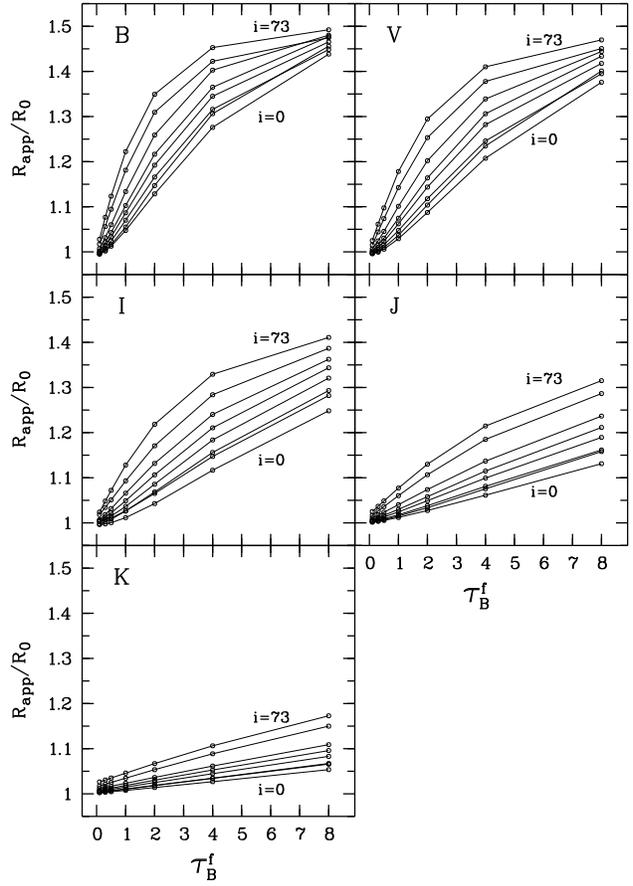,width=8.5cm,clip=}
\caption[]{The ratio of the apparent to intrinsic scalelengths 
$R_{\rm app}/R_{0}$ versus 
${\tau}^{f}_{\rm B}$, for $i=0, 26, 37, 46, 53, 60, 66, 73^{\circ}$,
and for the wavebands B, V, I, J and K.}
\label{figrtau5col}
\end{figure}
%================================================================
%=========== FIGURE: Rdn over inkl  =============
\begin{figure}
  \psfig{figure=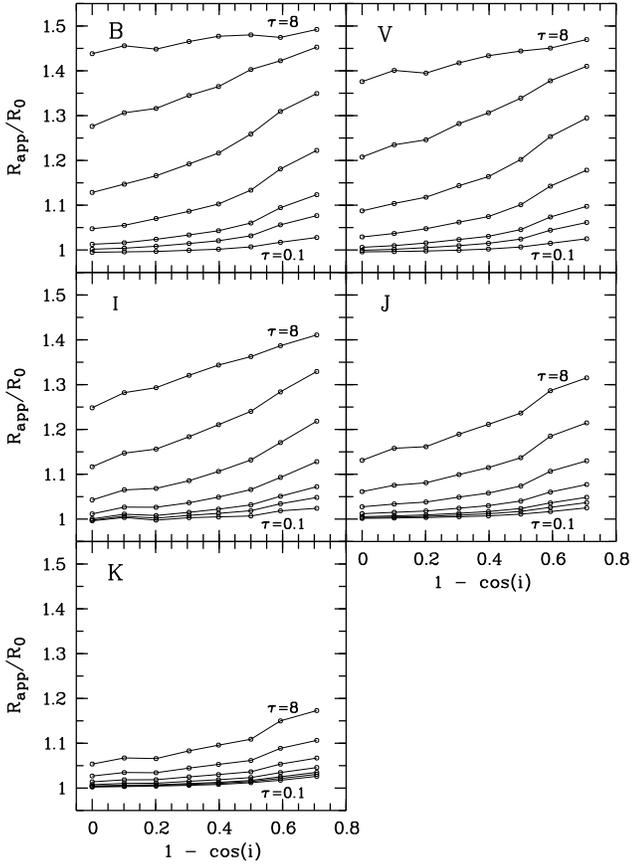,width=8.5cm,clip=}
\caption[]{The ratio of the apparent to intrinsic scalelengths 
$R_{\rm app}/R_{0}$ versus $1-cos(i)$, for 
${\tau}^{f}_{\rm B} = 0.1, 0.3, 0.5, 1.0, 2.0, 4.0, 8.0$, and for
the wavebands B, V, I, J, and K.}
\label{figrink5col}
\end{figure}
%================================================================

%=========== FIGURE: Rdn over color  =============
\begin{figure}
  \psfig{figure=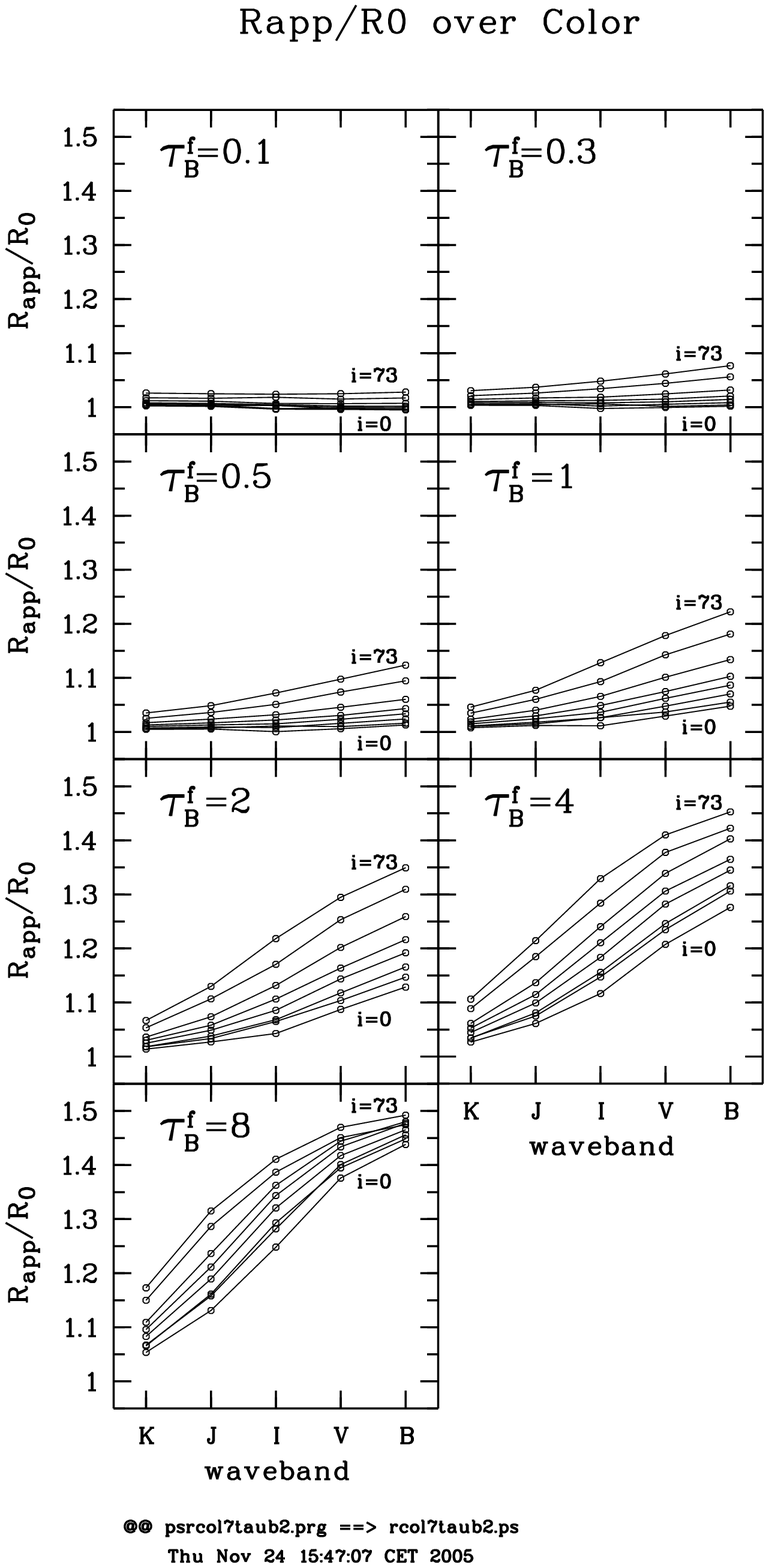,width=8.5cm,clip=}
\caption[]{The ratio of the apparent to intrinsic scalelengths 
$R_{\rm app}/R_{0}$ versus waveband, for  
 $i=0, 26, 37, 46, 53, 60, 66, 73^{\circ}$, and for 
${\tau}^{f}_{\rm B} = 0.1, 0.3, 0.5, 1.0, 2.0, 4.0, 8.0$}
\label{figrcol7tau}
\end{figure}
%================================================================
\subsection{Change of the exponential scalelengths}

In Fig.~3 we show the dependence of $R_{\rm app}/R_{0}$ on 
${\tau}^{f}_{\rm B}$, for different
inclinations and wavelengths. Because disks are always more attenuated in 
their central regions than in their outer regions, $R_{\rm app}/R_{0}$ is 
always greater than unity and increases monotonically with 
${\tau}^{f}_{\rm B}$. 
In some cases this increase is linear, whereas in others the increase flattens 
off towards higher values of ${\tau}^{f}_{\rm B}$. The linear behaviour 
happens when the disk is optically thin
along the lines of sight towards most of the projected radii. As
${\tau}^{f}_{\rm B}$ increases, an increasing area of the inner disk becomes
optically thick, producing a slowing-down in the increase with ${\tau}^{f}_{\rm
  B}$, until an asymptote is approached when the disk becomes optically thick
along the lines of sight over the whole extent of the disk. At longer
wavelengths, e.g. in the K band, such an asymptote is never reached and the
behaviour of the curves remains in the linear regime, since
even at the highest ${\tau}^{f}_{\rm B}$ and inclinations considered, the disk
remains optically thin along most of the lines of sight. By contrast, at the
shorter wavelengths, for instance in the B band, the asymptote is approached 
for high
${\tau}^{f}_{\rm B}$ and inclinations, as the disk becomes optically thick 
along most of the lines of sight. Furthermore, at these wavelengths the curves 
enter the non-linear regime at all inclinations, since some
part of the disk is always optically thick along the line of sight. At
intermediate wavelengths, such as in the I band, the variation with 
${\tau}^{f}_{\rm B}$ remains in the linear regime at lower inclinations, and 
only at high inclination is a non-linear behaviour seen, though without getting
near the asymptote. Another consequence of the existence of the 
asymptote 
is that the curves for different inclinations tend to bunch together at high 
${\tau}^{f}_{\rm B}$ and shorter wavelengths, as the asymptote is
approached. Thus, in B, V and I bands, the highest spread between the
curves at different inclinations is reached at intermediate ${\tau}^{f}_{\rm
  B}$, since, by necessity, the curves must also be bunched together at low 
${\tau}^{f}_{\rm B}$. In K band the curves never enter the non-linear
regime so the maximum spread happens at the highest ${\tau}^{f}_{\rm B}$.

In Fig.~4 we plot the same information as in Fig.~3, but as the 
variation of $R_{\rm app}/R_{0}$ with inclination, for different 
${\tau}^{f}_{\rm B}$ and wavelengths. The curves appear flat when the disk is
either predominantly optically thin at all inclinations (for example in K
band at ${\tau}^{f}_{\rm B}<2$, where $R_{\rm app}/R_{0}$ is close to unity) 
or when the disk is predominantly optically thick at all inclinations 
(for example in B band at ${\tau}^{f}_{\rm B}=8$, where $R_{\rm app}/R_{0}$ 
approaches the asymptote of $~1.5$).

In  Fig.~5 we plot $R_{\rm app}/R_{0}$ versus wavelength, for
different ${\tau}^{f}_{\rm B}$ and inclinations. The variation with wavelength
also shows a transition between a linear behaviour in an optically thin 
regime (for ${\tau}^{f}_{\rm B}<0.5$) and a non-linear approach to an 
asymptote at ${\tau}^{f}_{\rm B}=8$. 

In order to see what is the relative change in the scalelength between
different wavelengths just due to the effect of dust, one can look at  the
variation of the quantity
$\frac{\displaystyle R_{\rm app}(\lambda_1)}{\displaystyle 
R_{\rm app}(\lambda_2)} \times \frac{\displaystyle R_{0}(\lambda_2)}
{\displaystyle R_{0}(\lambda_1)}$ with  ${\tau}^{f}_{\rm B}$. As an 
example, we plot in Fig.~6 this quantity for the wavelengths B and I. At all
inclinations, the curves show a similar basic shape. The curves rise steeply at
low  ${\tau}^{f}_{\rm B}$, reach a maximum and then decline slowly. This
behaviour is compressed over a smaller range of ${\tau}^{f}_{\rm B}$ for higher
inclinations. Because of this,  
$\frac{\displaystyle R_{\rm app}(B)}{\displaystyle 
R_{\rm app}(I)} \times \frac{\displaystyle R_{0}(I)}
{\displaystyle R_{0}(B)}$ increases with inclination at low 
${\tau}^{f}_{\rm B}$, but decreases with inclination at high ${\tau}^{f}_{\rm
  B}$, with the curves crossing at intermediate ${\tau}^{f}_{\rm B}$. In other
words, for a low ${\tau}^{f}_{\rm B}$, in order to convert the ratio between
the apparent scalelength in B and I to the corresponding intrinsic ratio, one
needs to apply a correction factor which is larger for high inclinations than
for low inclinations, whereas at high  ${\tau}^{f}_{\rm B}$ the opposite
is true. An interesting feature of these curves is that they are relatively 
flat for ${\tau}^{f}_{\rm B}>2$.
%which means that, provided a galaxy is
%not too optically thin, the ratio of its apparent scalelengths at B and I can 
%be converted to the corresponding intrinsic ratio to an accuracy of a
%few percent, even if the exact value of ${\tau}^{f}_{\rm B}$ is not known.
%=========== FIGURE: Rd(B)/Rd(I)  only dust =============
\begin{figure}
  \psfig{figure=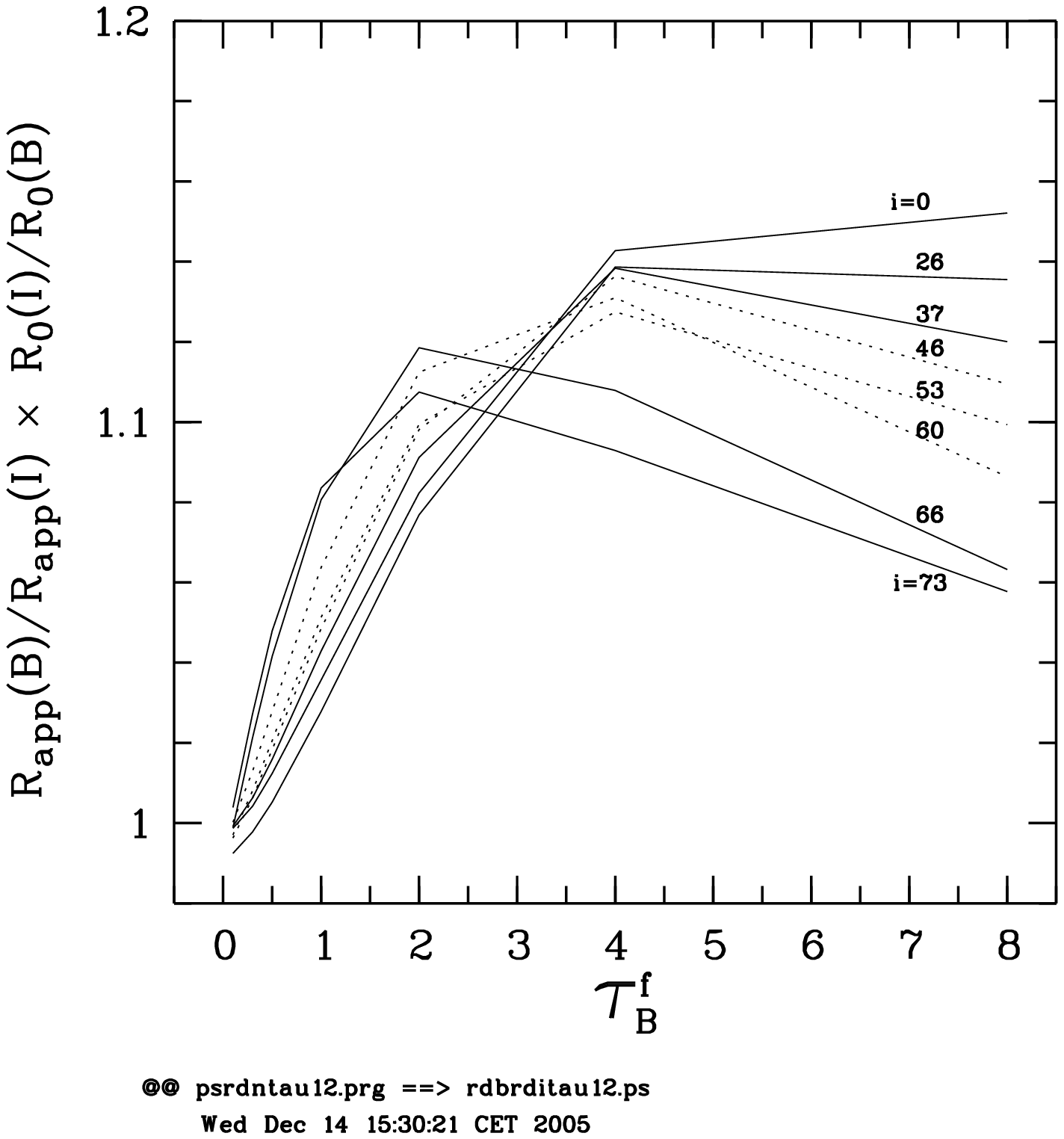,width=8.5cm,clip=}
\caption[]{The change in the ratio of the scalelengths in B and I bands due
to the effects of dust, plotted as 
$\frac{\displaystyle R_{\rm app}(B)}{\displaystyle 
R_{\rm app}(I)} \times \frac{\displaystyle R_{0}(I)}{\displaystyle R_{0}(B)}$
versus ${\tau}^{f}_{\rm B}$.
Different curves are plotted for a range of inclinations from $i=0$ to
$i=73$. The curves for intermediate inclinations ($i=46$, $i=53$ and $i=60$)
are plotted with dotted lines for clarity.}  
\label{figrdbrdi}
\end{figure}

\begin{figure}
 \psfig{figure=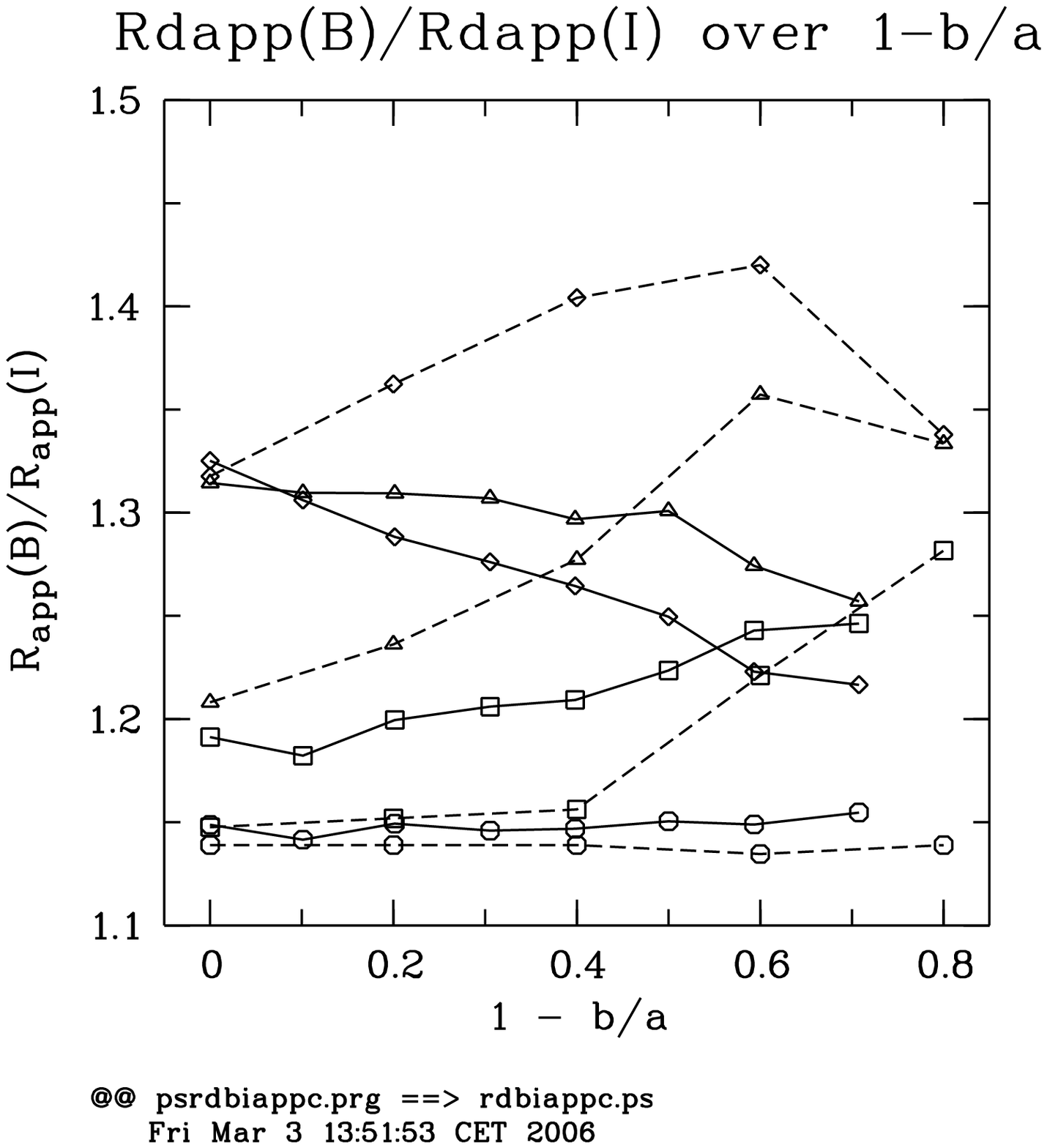,width=8.5cm,clip=}
\caption[]{Comparison between the apparent ratios of scalelength in B and I
  predicted by our work (solid lines) and by the work of Cunow et al. (2001)
  (dashed lines). 
The curves are for ${\tau}^{f}_{\rm B}$ = 0.1, 1, 4 and 8 (interpolated in 
the case of the Cunow curves). The pairwise corresponding curves are marked 
by 4 different symbols.}
\end{figure}
 
Simulations of the observed ratios of scalelength in B and I as a function of
inclination and ${\tau}^{f}_{\rm B}$ have also been performed by Cunow et 
al. (2001). The closest comparison possible is that with model (k) of Cunow, 
which differs from our model only through the absence of the dust disk 
associated with the young stellar population. In Fig.~7 we have plotted the
ratios of apparent scalelength in B and I as a function of apparent ellipticity
$(1-b/a)$ for both our model and the model from Fig.~7 of Cunow et al. (2001),
where the ratios from Cunow have been interpolated to match the opacities
given in our tables. Overall the curves of Cunow behave as if their model is
more optically thin for the same ${\tau}^{f}_{\rm B}$, which is the opposite
behaviour one would expect from the differences in the geometrical
distributions between the models. However the procedure to derive scalelength
followed by Cunow et al. only uses data from the outer parts of the disk having
a linear profile, whereas our method uses data from the whole disk. This
obviously biases the results from Cunow towards more optically thin solutions
and this effect dominates over the differences induced by the different 
geometrical distributions. 

Another work on predicting apparent scalelength ratios between B and I is that
of Byun et al. (1994). These results are very similar to those of Cunow et
al. (2001) (see Fig.~8 of Cunow et al.), which may reflect the fact that
Byun et al. also use extrapolation from the outer disk in deriving photometric
parameters. 

\subsection{Change of the central surface brightness}

In Fig.~8 we show the dependence of $\Delta$SB on ${\tau}^{f}_{\rm B}$, for 
different inclinations and wavelengths. 
An interesting feature of these curves is that they pass through both positive
and negative values of $\Delta$SB: the centre of the disk can appear either 
dimmer or brighter than the face-on dustless disk. This is because the
projected surface density of stars visible along the line of sight can be 
either greater or smaller than the surface density of stars seen towards the 
centre of a face-on dustless disk. The projected surface density of stars can
be increased by viewing the disk at higher inclinations and can be decreased by
attenuating the stellar light with dust. In the K band the central part of the
disk appears brighter than the centre of the face-on dustless disk at 
almost all inclinations and ${\tau}^{f}_{\rm B}$, because the line of sight
through the centre is almost always transparent. On the contrary, in the B band
the opposite is true, since the line of sight through the centre is almost
always opaque. 

\begin{figure}
  \psfig{figure=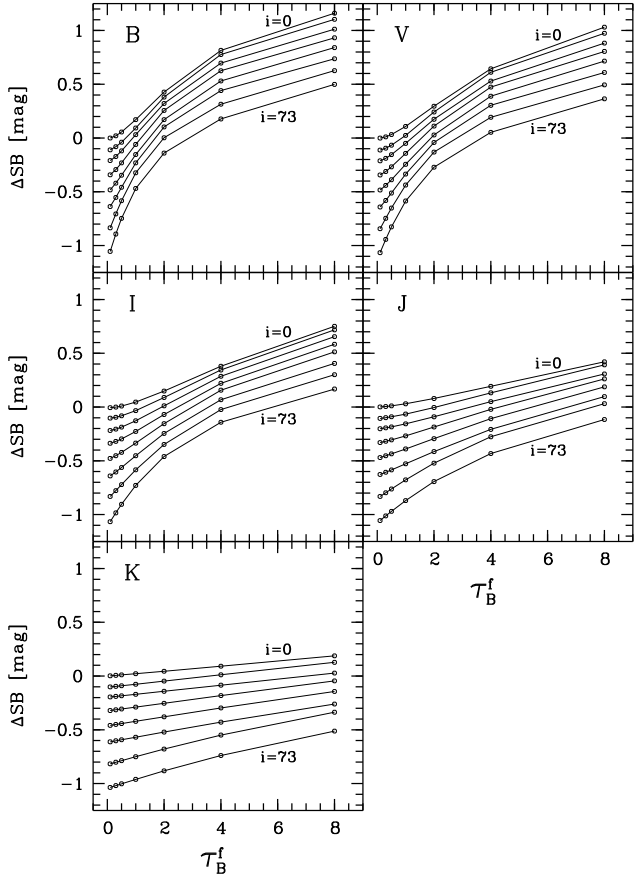,width=8.5cm,clip=}
\caption[]{The ratio of the apparent average central surface brightness to 
the intrinsic average central face-on surface brightness expressed in
magnitudes, $\Delta\rm{SB}$,  
versus ${\tau}^{f}_{\rm B}$, for $i=0, 26, 37, 46, 53, 60, 66, 73^{\circ}$,
and for the wavebands B, V, I, J, and K.} 
\label{figdsb5col}
\end{figure}

\begin{figure}
  \psfig{figure=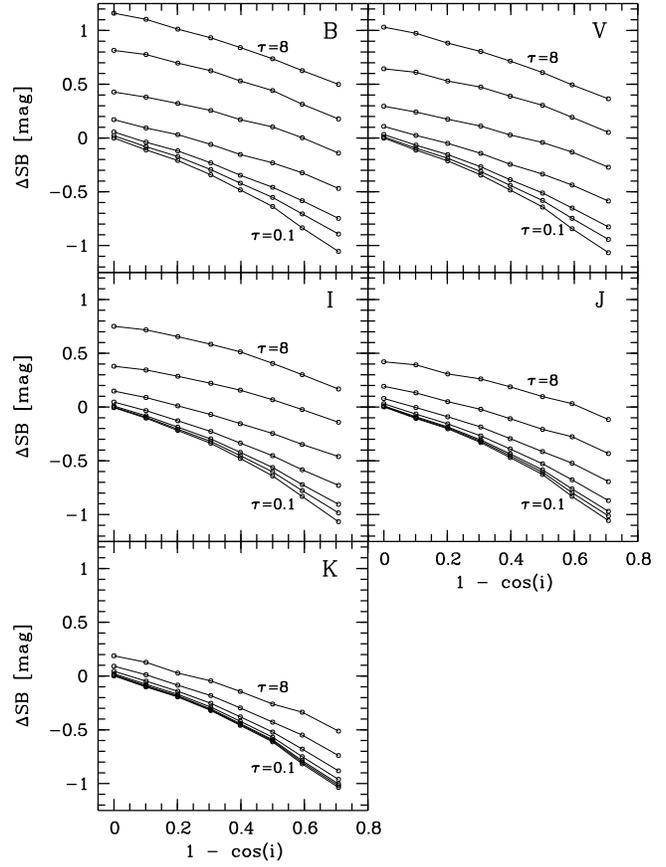,width=8.5cm,clip=}
\caption[]{The ratio of the apparent average central surface brightness to 
the intrinsic average central face-on surface brightness expressed in
magnitudes, $\Delta\rm{SB}$, versus $1-cos(i)$, for ${\tau}^{f}_{\rm B}=0.1, 
0.3, 0.5, 1.0, 2.0, 4.0, 8.0$ and for the wavebands B, V, I, J, and K.} 
\label{figdsb5col}
\end{figure}

\begin{figure}
  \psfig{figure=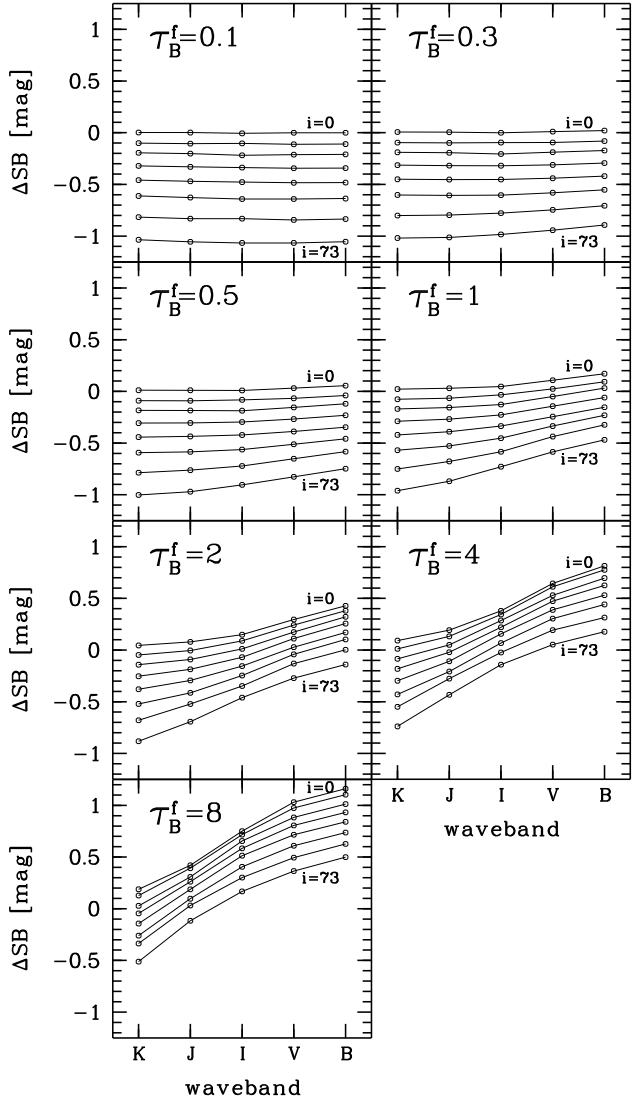,width=8.5cm,clip=}
\caption[]{The ratio of the apparent average central surface brightness to 
the intrinsic average central face-on surface brightness expressed in
magnitudes, $\Delta\rm{SB}$, versus waveband, for 
$i=0, 26, 37, 46, 53, 60, 66, 73^{\circ}$,
and for ${\tau}^{f}_{\rm B}=0.1, 0.3, 0.5, 1.0, 2.0, 4.0, 8.0$.} 
\label{figdsb5col}
\end{figure}

The form of the curves in Fig.~8 can be explained in terms of the two 
competing factors viewing angle and opacity:
the brightening of the  $SB_{\rm app}^{\rm c}$ with increasing inclination 
and the dimming of the $SB_{\rm app}^{\rm c}$ with increasing 
${\tau}^{f}_{\rm B}$. 
The former effect is
independent of wavelength, since the line of sight of stars is always the same
for a certain inclination\footnote{Formally there is a very small dependence on
  wavelength due to the fact that the scalelength of the stars varies with
  wavelength in our model. In practice, however, this effect is negligible over
  the range of inclinations considered here and cannot be seen in the plot.}
%The independence on wavelength would no
%  longer hold if the disk is viewed closer to edge-on, since then the column
%  density of stars would depend on the scalelength of
%  the stars, and this scalelength varies, in our model, with wavelength.}. 
This can be seen 
from the wavelength independence of the spread in $\Delta$SB from 0 to -1 
magnitude for ${\tau}^{f}_{\rm B}=0$. The latter effect (the dimming due to
opacity) is strongly dependent
on wavelength, since,  for a certain ${\tau}^{f}_{\rm B}$, the depth into 
which stars can be seen along the line of sight decreases with decreasing 
wavelength. The dimming of  $SB_{\rm app}^{\rm c}$  with increasing
${\tau}^{f}_{\rm B}$ is almost negligible in the K band at inclinations
close to face-on, when the line of sight through the centre is almost
transparent.  On the contrary, in the B band this effect becomes strong at all inclinations.

\begin{figure}
 \psfig{figure=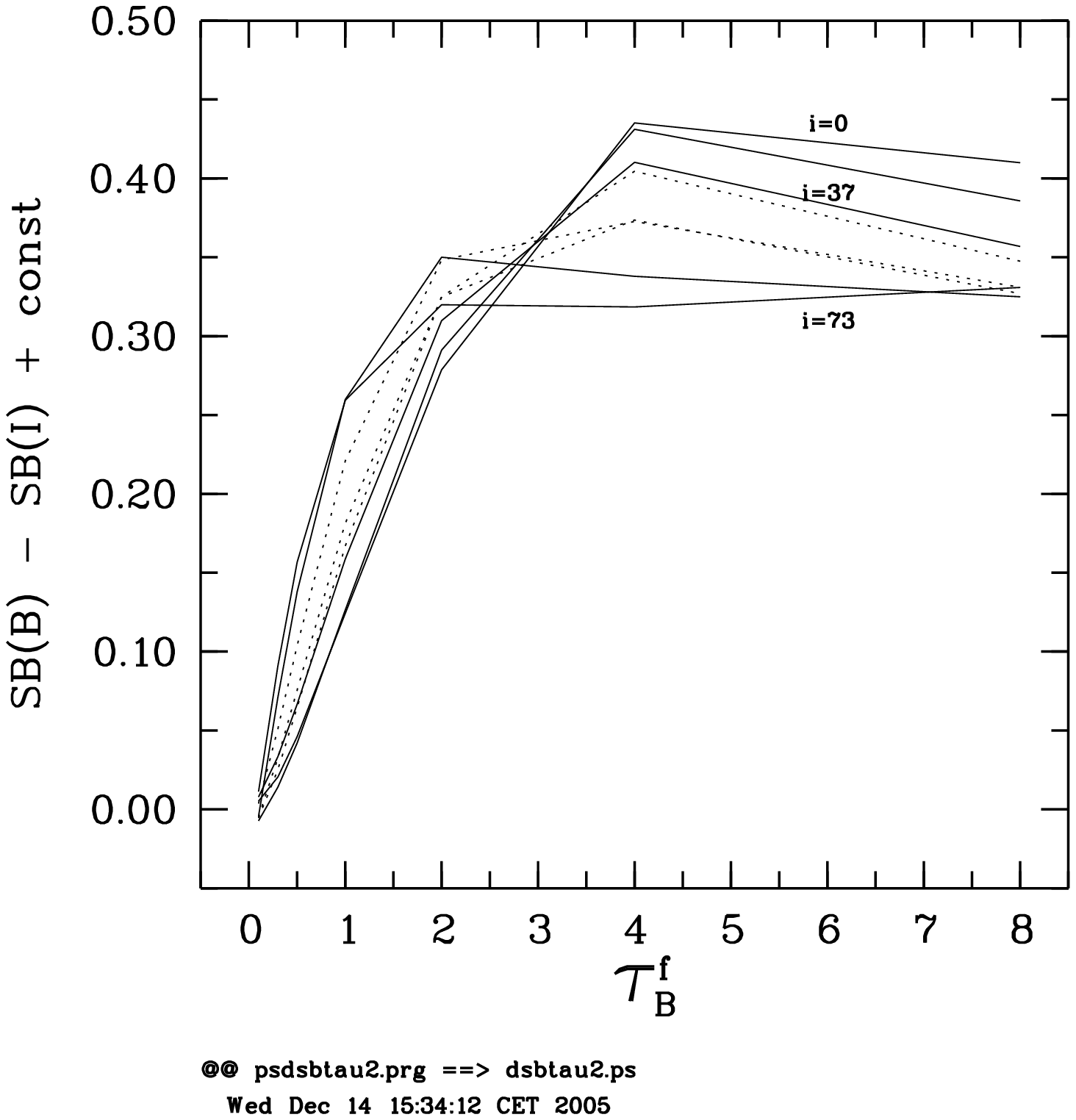,width=8.5cm,clip=}
\caption[]{The change in the B-I central colour due to the effects of dust, 
plotted as $\Delta SB(B)-\Delta SB(I)+ const$ versus ${\tau}^{f}_{\rm B}$. 
Different curves are plotted for a range of inclinations from $i=0$ to
$i=73$. The curves for intermediate inclinations ($i=46$, $i=53$ and $i=60$)
are plotted with dotted lines for clarity.}
\end{figure}

Another feature of these curves is the transition between a linear dependence
of  $\Delta$SB on ${\tau}^{f}_{\rm B}$ when the lines of sight through the
centre are optically thin (e.g. in the K band at all inclinations), to a
non-linear regime for optically thick lines of sight (e.g. in the B band at 
all
inclinations). At intermediate wavelengths (e.g. in the J band) one sees a
transition between these two regimes: the upper curves (close to face-on) are
in the linear regime, whereas the lower curves (high inclinations) are
non-linear. 

For a given (non-zero) ${\tau}^{f}_{\rm B}$, the dispersion between the 
curves in Fig.~8 for low and high inclinations varies systematically with
wavelength. For the case of  ${\tau}^{f}_{\rm B}=4$ one can see that the
dispersion has a maximum in the K band, then it decreases towards I band, after
which it increases again towards B band. At first glance this behaviour is not
very intuitive. If the stars and dust had an 
identical distribution, then one
would expect a monotonic decrease in dispersion when moving from
transparent central disks to opaque central disks, until the lines of
sight would become completely opaque at all inclinations and the 
dispersion would tend to zero (since the brightness of an optically thick 
body is independent of the viewing angle). The fact that the behaviour is 
more complex than this is a consequence of the non-identical distribution of 
stars and dust in the vertical direction. In our model the stars and
dust are distributed in exponential disks, with the stars having
scaleheights larger than those of the dust. As already remarked by Disney et
al. (1989), ``even in a highly obscured galaxy, the upper layers of unobscured
stars will apparently increase its mean surface brightness as it is
inclined, mimicking the behaviour of an optically thin system''. Thus, at the 
extreme of very high 
optical depth one approaches the same dispersion as for the very low optical
depth, 
since the variation in the surface brightness between low and high inclination
is only produced by the stars above the (exponential) dust layer.
At intermediate optical depths the dispersion over inclination is lower, 
because a larger fraction of this dispersion comes from stars well mixed
with the dust which have a smaller variation in the surface brightness 
between low and high inclination. In our model, there will always be an
increase in the surface brightness with increasing inclination, even for
optically thick disks. This can be seen even better in Fig.~9, where we plot 
the same information as in Fig.~8, but as the variation of  $\Delta$SB with 
inclination, for different ${\tau}^{f}_{\rm B}$ and wavelengths. We also
plotted the variation $\Delta$SB with wavelength (Fig.~10), for 
different ${\tau}^{f}_{\rm B}$ and inclinations, to better capture the
transition between the almost independence of $\Delta$SB on wavelength at 
low ${\tau}^{f}_{\rm B}$ to a strong dependence on wavelength at high
${\tau}^{f}_{\rm B}$.

In order to see what is the change in the central colour of the disk due to
dust, one can look at the variation of
$\Delta SB(\lambda_{1})-\Delta SB(\lambda_{2})$ with  ${\tau}^{f}_{\rm
  B}$. As for the analogous plot for the variation of scalelength, we plot in
Fig.~11 this quantity for the wavebands B and I, $\Delta SB(B)-\Delta SB(I)$,
which is equivalent to a plot of SB(B) - SB(I) shifted by a constant value,
such that the curves go through the origin. These curves, therefore, give the
colour correction that needs to be applied to correct apparent colour to
intrinsic colour at the centre of the disk. The curves show the same
qualitative behaviour as those for the relative change in scalelength, namely
that they rise steeply at
low  ${\tau}^{f}_{\rm B}$, reach a maximum and then decline slowly.
We note again the relatively flat behaviour of the curves for 
${\tau}^{f}_{\rm B}>2$. 
%means 
%that, provided a galaxy is not too optically thin, the apparent B-I central 
%colour can be converted to the corresponding intrinsic colour to an accuracy 
%of about $\pm 0.07$ mag, even if the exact value of ${\tau}^{f}_{\rm B}$ is 
%not known.

%%%%%%%%%%%%%%%%%%%%%%%%%%%%%%%%%%%%%%%%%%%%%%%%%%%%%%%%%%%%%%%%%%%

\subsection{Change of the axis ratios}
\label{changeqd}

In Fig.~12 we show the dependence of the ratio of the apparent to the intrinsic
axis ratio, $Q_{app}$/$Q_{0}$ on ${\tau}^{f}_{\rm B}$, for different
inclinations and wavelengths. One can see that this ratio can be either greater
or less than one, which means that the galaxy can appear less or more inclined
than it is in reality. In other words the galaxy can appear rounder or 
flatter than the shape of an infinitely thin dustless disk at the true 
inclination of the galaxy.
This behaviour is attributed to two effects that act in the opposite sense. 
Firstly, the vertical distribution of stars makes the galaxy appear rounder 
than an infinitely thin dustless disk. Secondly,
the presence of dust makes the galaxy appear flatter than an infinitely thin
dustless disk. The first effect is obvious, but the second needs some
explaining. As one increases the line of sight optical depth, either by
increasing the inclination or the ${\tau}^{f}_{\rm B}$, the
observed light becomes dominated by photons originating
from an increasingly thin stellar layer on the side of the dust disk nearest to
the observer, due to absorption by dust of photons originating
from below this layer.

The effect of the vertical distribution of stars is best seen by looking at the
inclination dependence of $Q_{app}$/$Q_{0}$ for ${\tau}^{f}_{\rm B}$ close to
zero. For example in the K band, $Q_{app}$/$Q_{0}$ increases monotonically with
inclination up to $\sim 1.13$ for $i=73^{\circ}$. 
%This value is determined by
%the ratio of the scaleheight to scalelength of the intrinsic stellar
%distribution. Because in our model this ratio decreases with decreasing
%wavelength, the value of  $Q_{app}$/$Q_{0}$ at $i=73$ degrees 
%and ${\tau}^{f}_{\rm B}$ close to zero decreases in
%progressing from K to B band. In other words dustless galaxies will appear
%flatter in the B band than in the K band at large inclination.

The effect of opacity is more complex. One feature is the decrease of 
$Q_{app}$/$Q_{0}$ with increasing ${\tau}^{f}_{\rm B}$, for all inclinations
and wavelengths. This effect is weak in the K band where opacity is small, and
is strong in the B band where opacity is large. A further aspect
of the effect of dust is that at high opacity, and except for inclinations 
near to edge-on, the dependence of $Q_{app}$/$Q_{0}$ on inclination is
the opposite of that predicted by the model at low opacity. Thus, in B-band at
${\tau}^{f}_{\rm B}=8$, the general trend is for $Q_{app}$/$Q_{0}$ to 
decrease with increasing inclination\footnote{within the numerical noise
  induced by the finite numbers of directions for which scattered light is
  calculated.} over the range $i=0$ to $i=66$,
as one progressively sees a
thinner and thinner layer of stars above the dust layer, making the galaxy
appear flatter with increasing inclination. 
Suddenly, at $i=73$,
this trend is reversed, as one starts to see stars both above and below the 
dust layer, and a big step is made to a rounder shape. The different behaviour
of $Q_{app}$/$Q_{0}$ on inclination at low and high ${\tau}^{f}_{\rm B}=8$ 
means that the curves cross at intermediate  ${\tau}^{f}_{\rm B}$.

%%%%%%%%%%%%%%%%%%%%%%%%%%%%%%%%%%%%%%%%%%%%%%%%%%%%%%%%%%%%%%%%%%%%
\begin{figure}
 \psfig{figure=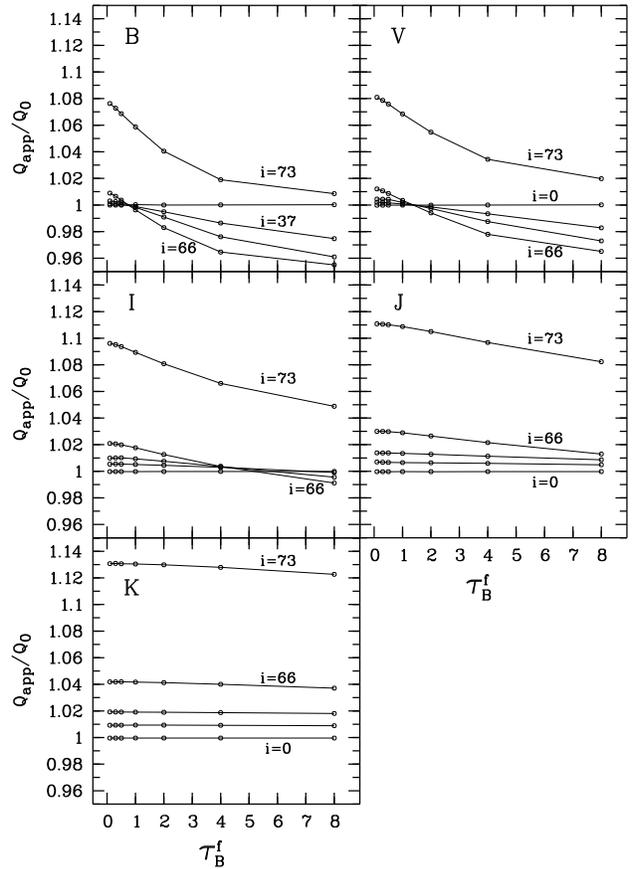,width=8.5cm,clip=}
\caption[]{The ratio of the apparent to intrinsic axis ratio  $Q_{app}$/$Q_{0}$
versus ${\tau}^{f}_{\rm B}$, for $i=0, 37, 53, 66, 73^{\circ}$ and for the
wavebands B, V, I, J, and K.}
\end{figure}

%%%%%%%%%%%%%%%%%%%%%%%%%%%%%%%%%%%%%%%%%%%%%%%%%%%%%%%%%%%%%%%%%%%%
%%%%%%%%%%%%%%%%%%%%%%%%%%%%%%%%%%%%%%%%%%%%%%%%%%%%%%%%%%%%%%%%%%%%
\section{Discussion}

Having established the dependence on inclination, face-on opacity and
wavelength of the corrections needed to convert the apparent
photometric parameters scalelength, central surface brightness and axis 
ratios into the corresponding intrinsic quantities, we can now discuss the
implications of these findings. But before doing this, we 
caution that the corrections given in this
paper are valid only for disk-dominated galaxies. In reality, many galaxies
have prominent bulges and therefore the apparent photometric parameters of the disk 
(including scalelength, central surface brightness and total magnitudes) have
to be found in a simultaneous fit of geometrical functions for the bulge and 
the disk. However, the apparent photometric parameters derived from this 
bulge-to-disk
decomposition will be a function of both opacity and bulge-to-disk ratio. 
In other words, the apparent scalelength and central surface brightness that 
one derives for a disk galaxy
with a given opacity and inclination in the presence of a bulge will in
general be different from
the apparent scalelength that one derives for the same disk (with the same
opacity and inclination) without a bulge. Therefore one should restrict the
analysis to galaxies having small bulges, where such effects can be ignored.

One should also caution that the corrections for dust given in
  this paper are calculated assuming a small wavelength dependence of the ratio
  of the intrinsic scalelength of the stars to that of the dust, as specified
  in Tuffs et al. (2004). Thus, strictly speaking, the tabulated corrections 
are valid only for this geometry. In practice, however, these corrections can
  be used also for systems having a different variation in the ratio of the 
intrinsic scalelengths. Comparison
  between Fig.~7b and h from Cunow et al. (2001) show that, if one were to make
  an error of $25\%$ in the assumed intrinsic scalelength of stars relative to 
  the dust, the corresponding error in the effect of dust would be at most a 
few percent for typical opacities and inclinations less than $70^{\circ}$. 
Only
  for ${\tau}^{f}_{\rm B} > 7$ and inclination greater than $70^{\circ}$ 
does the error exceed
  $10\%$.

After these words of caution we can now examine the
impact of these new results on our ability to derive quantities of 
astrophysical interest from optical observations of disk-dominated galaxies.
In principle, if one knew 
the face-on opacity, one could use the correction factors to analyse
the intrinsic properties of galaxies on an object to object basis. In practice,
however, the face-on opacity of an individual galaxy can only be reliably 
determined by self-consistently modelling its combined UV/optical/FIR/submm 
output, which is at present only 
rarely possible, since the vast majority of galaxies has not yet been
 measured  
in the FIR/submm regime. For large statistical samples, though, one can derive 
a characteristic value for the face-on opacity of a population of systems from
the optical measurements themselves by 
analysing the inclination dependence of the photometric parameters described 
here. 

One can also use our tabulated corrections for the investigation of 
astrophysical effects for which, although the absolute effect of dust is
non-negligible,
the predicted variation over a likely range in opacity is relatively small, 
such that
an exact knowledge of opacity is not needed. An example is the variation of the
scalelength between B and I due to dust, which is small, provided
 that ${\tau}^{f}_{\rm B}> 2$ (galaxies are not too
optically thin).\footnote{In our detailed SED modelling of
individual galaxies given in Paper ~I and II we showed that this range in
opacity is quite likely to characterise spiral galaxies.} 
Indeed, inspection of Fig.~6 shows that for a galaxy at a typical 
inclination of $37^{\circ}$ and having any ${\tau}^{f}_{\rm B}>2$, the effect of dust is to increase the
scalelength in B relative to that in I by a factor of $1.12 \pm 0.02$. The
scatter in this ratio is so small that one can compare the intrinsic 
scalelength of the stellar populations in B and I by simply applying a 
single correction factor for dust for any spiral galaxy at this
inclination. Similarly, other correction factors can be derived from our tables
for other inclinations and other waveband combinations.

Correspondingly, one can also compare the surface brightnesses between B and 
I to get
information about the intrinsic colour of the central and most dust affected
regions of the disks, even without detailed knowledge of 
${\tau}^{f}_{\rm B}$.  Inspection of Fig.~11 shows that for a galaxy at an
inclination of $37^{\circ}$ and for ${\tau}^{f}_{\rm B}>2$, the change in B-I
central colour due to dust is $0.36\pm 0.05$ magnitudes. Again, the scatter in
the central colour is small enough that one can look at the intrinsic central
colour between B and I by simply applying a single correction factor for dust
for any spiral galaxy at this inclination.

Lack of knowledge of opacity may be somewhat more problematic when one 
attempts to compare the properties of dusty galaxies in the local and distant 
universe at a fixed restwavelength.
When studying the evolution of disk sizes with redshift (e.g. by comparing
the scalelength of low ($z\sim 0$) and high ($z\sim 1$) redshifted galaxies 
at a common restwavelength)
one must correct for the increase in the apparent disk scalelength with
increasing redshift, stemming from the expected increase in opacity.
If, for example,  the opacity increased from 4 at $z=0$ to 8 or more at 
$z=1$, the apparent scalelength in the B band  of a galaxy seen at $i=37^{\circ}$  would increase by a factor of at least 1.10 due to dust.
This factor would rise to at least 1.24 for an increase in opacity from 2 
to 8 or more. The increase in the apparent scalelength with increasing 
redshift due to the
effect of dust goes in the opposite direction to the
predicted evolution in intrinsic disk size, and in general would need to 
be taken into account in quantitative analyses. Ultimately one would 
need to refine the technique to
take into account the possible evolution of the dust/star geometry\footnote{
  For
  example, depending on the assumed star-formation history, it may be 
  necessary to
  invoke different star/dust geometries to explain the observed colours of
  distant disk galaxies (Pierini et al. 2005).} when making
corrections for high-redshifted galaxies. This will become possible when
spatially resolved dust emission data from the rest-frame far-infrared is 
available for such systems, thereby permitting an analysis of the 
star/dust geometry to be performed analogous to that performed for local 
universe galaxies.

Our tabulated corrections for dust will also allow the analysis of the
correlation between the central face-on surface brightness versus scalelength,
as investigated by Graham (2001) in the K band where the galaxies are optically
thin, to be extended to the shorter wavebands where the galaxies may be 
optically thick. As a by-product of such an investigation, we note that one
could obtain a statistical determination of ${\tau}^{f}_{\rm B}$ from the
inclination dependence of both central surface-brightness and scalelength. 
As previously cautioned, however, this analysis has for the moment to be
restricted to disk-dominated galaxies.

%================================================================
\section{Application to local universe disk galaxies}

\begin{figure}
\psfig{figure=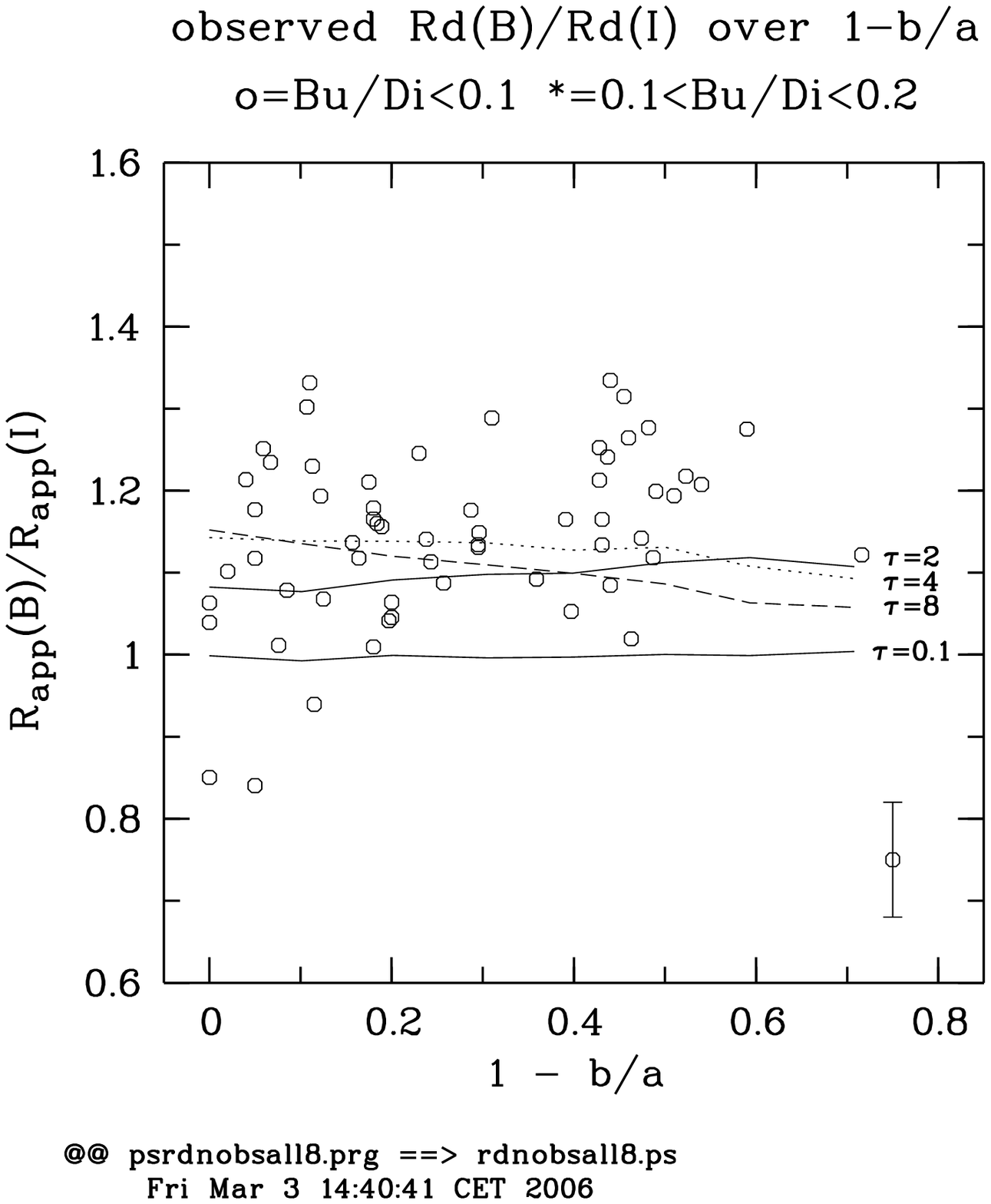,width=8.5cm,clip=}
\caption[]{Observed scalelength ratios between B and I, $\frac{\displaystyle R_{\rm app}(B)}{\displaystyle 
R_{\rm app}(I)}$, versus observed axis ratios $1-b/a$. The lines represent the
model predictions for the same relation due only to the effect of dust, which
is equivalent to the prediction of apparent scalelength ratios for dusty disks 
having the same intrinsic scalelength in B and I. The lines correspond to
${\tau}^{f}_{\rm  B}=0.1,2,4,8$. A typical error bar of $7\%$ is plotted in
the lower right hand corner.}
%for dusty disks with the same intrinsic scalelengths at 
%B and I with  ${\tau}^{f}_{\rm B}=0.1,2,4,8$.
\end{figure}

As mentioned in the introduction, one of the motivations of this work is to
quantify the change in the intrinsic scalelength of different stellar
populations in local universe disk galaxies. Here we give an example of the
application of our model to such an
investigation. For this we used published apparent disk scalelengths in 
B and I, restricting the data to galaxies which have measured bulge-to-disk 
ratios of less than 0.2. There are 
two  reasons for this selection criterion. Firstly, as mentioned in the 
discussion, one needs to restrict the analysis to disks with small
bulges, since the apparent scalelength derived for dusty disks with large 
bulges will diverge from the apparent scalelength of pure
dusty disks.
Secondly, for systems with small bulges MacArthur et al. (2003) showed that 
results from one-dimensional 
photometry are equivalent to those derived from two-dimensional fits (as 
required by this work). The
resulting sample is comprised of 61 galaxies taken
from M\"ollenhoff (2004), MacArthur et al. (2003),  Prieto et al. (2001) and  
Pompei \& Natali (1997). Since MacArthur et al. had measurements in B V R H, 
their R band scalelengths were converted to I band scalelengths 
by multiplying them by a factor of 0.981. This factor was obtained by linearly
interpolating between the average scalelength in R and H, as given in Table~3
of MacArthur et al. . 
 Despite our selection of disk-dominated galaxies and the known tendency 
of such
systems to preferentially populate the intermediate- and low-mass end of the 
mass distribution of nearby, late-type galaxies (e.g. Pierini \& Tuffs 1999), 
the galaxies used in this sample are all bright spiral galaxies with absolute
magnitudes ${M_{\rm V}}$ = -19.2 to -22.3. Thus they have geometries of stars 
and dust appropriate to giant spirals, which are well described by our model.

In Fig.~13 we plot observed scalelength ratios between B and I, $\frac{\displaystyle R_{\rm app}(B)}{\displaystyle 
R_{\rm app}(I)}$, versus axis ratios $1-b/a$.
All but three of the 61 data points lie above unity, which means that almost
all galaxies have apparent scalelengths which are larger in B than in I. The 
data also show a scatter, which is consistent with the
random errors estimated for the measurements.\footnote{ 
The errors of a single scalelength determination 
were estimated to be $\sim 5\%$ in M\"ollenhoff, $\sim 3\%$ in Cunow and 
 $\sim 5\%$ in MacArthur et al.. Thus the error of the quotients in 
Fig.~13 can be estimated to be $\sim 7\%$. A corresponding error bar is 
   plotted there.}
The mean and the standard deviation in $R_{\rm app}(B)/R_{\rm
app}(I)$ is 1.156 and 0.094, respectively.
In order to see whether these observed values can be accounted
for by the effect of dust alone, we compare them with model predictions for
${\tau}^{f}_{\rm B}>2$ (see Fig.~13) which is a reasonable range in ${\tau}^{f}_{\rm B}$ to expect for 
disk galaxies. 
For each of the model predictions we calculated the corresponding ratio of intrinsic scalelengths at B and I
which best fit the data by shifting the lines vertically in the figure
until an equal number of data points appeared above and below the line.
This process yielded 
average ratios of intrinsic scalelengths
 of 1.057, 1.019 and 1.037, for ${\tau}^{f}_{\rm B}=2, 4, 8$, respectively, 
and an overall range ($\pm 1\sigma$ for the sample) of intrinsic scalelength
ratios from 0.93 to 1.15. Because the deviation from unity of the averaged 
ratios is so much less than the spread in the individual data points, we
conclude that, on average, and within the available
statistics,
there is almost no systematic increase in the intrinsic scalelength
with decreasing wavelength and that the observed predominance of
galaxies with ratios of apparent scalelengths greater than unity is mainly
due to the effects of dust. The data would only admit a small 
variation of the intrinsic stellar scalelength between B and I (less than 1.15)
if ${\tau}^{f}_{\rm  B}<2$. However combined analysis of the UV/optical and 
FIR/submm have generally yielded ${\tau}^{f}_{\rm B}<2$ 
(Bianchi et al. 2000b,  Paper I, II, Misiriotis et al. 2004, Meijerink et al. 2005).

Much bigger statistical samples will be needed
to measure systematic trends in intrinsic scalelengths with wavelength,
if they exist. Finally we note that the scatter in the observed ratio of 
scalelengths between B and I (with $\sigma=0.094$) is comparable to the
estimated errors in this quantity for individual galaxies and can therefore 
primarily be ascribed to measurement errors.
%is about twice the maximum conceivable scatter which can 
%be attributed to the effects of dust alone. {\bf Formally, the scatter thus 
%indicates
%real differences in the intrinsic scalelength of the stellar populations
%sampled at B and I, though in practice the scatter could also be due to unforseen 
%measurement errors.}

Our conclusion that dust is mainly responsible for the larger scalelength
  observed in the B band compared to I band can be contrasted with the results
  from Cunow (2001, 2004), who deduced that dust alone was not sufficient to
  explain this effect. Here we note that the different conclusions can 
  primarily be attributed to differences in the data rather than differences 
  in the model. In fact,
  as noted by Cunow (2004), her data for normal galaxies, both in terms of the
  absolute values of the scalelength ratios and the variation of these ratios
  with inclination, cannot be fitted by any model. As shown in Fig.~13, our 
  data is reasonably well fitted by our model in both these respects.

Another study in the literature that has addressed the wavelength dependence of
scalelength (this time between B and K bands) is that of Peletier et
al. (1994). Here again, on average, the scalelength ratios have much larger 
values than in our sample, even taken into account the different wavelength
used. Although Peletier et al. interpreted their results as being due to the
effect of dust on the apparent scalelengths, such large effects are not
predicted by any dust model. Bearing in mind the inclusion of galaxies with
bulges in the sample of Peletier et al., such effects are, in our view, more 
likely to be produced by the effect of dust on the bulge-disk decomposition 
itself. This is because the use of simple templates for dustless bulge and 
disk in bulge-disk decomposition 
would lead to systematic bias in the derived photometric parameters, 
including apparent disk scalelength.
 
%%%%%%%%%%%%%%%%%%%%%%%%%%%%%%%%%%%%%%%%%%%%%%%%%%%%%%%%%%%%%%%%%%%%

%%%%%%%%%%%%%%%%%%%%%%%%%%%%%%%%%%%%%%%%%%%%%%%%%%%%%%%%%%%%%%%%%%%%
%%%%%%%%%%%%%%%%%%%%%%%%%%%%%%%%%%%%%%%%%%%%%%%%%%%%%%%%%%%%%%%%%%%%
\section{Conclusions and summary}
\label{conclusion}
We have fitted the simulated images of dusty disk galaxies presented in 
Paper~III with a template brightness distribution corresponding to an 
inclined infinitely thin dustless exponential disk to obtain apparent 
scalelengths, central surface-brightness distributions and axis ratios. These
are the apparent photometric quantities that an observer would extract from 
observed images of galaxies. Using the prior knowledge of the corresponding 
intrinsic quantities input in the simulations, we were able to derive the
correction factors (listed in Tables~1--5 of this paper) for the conversion 
of the apparent to intrinsic quantities due to dust, as a function of 
inclination $i$, central face-on optical depth in B band  ${\tau}^{f}_{\rm B}$,
and wavelength.

The apparent to intrinsic scalelength ratio is always greater than unity and 
can vary up to 50$\%$. The apparent to intrinsic central surface-brightness
ratio expressed in magnitudes can be either positive or negative, depending on
whether the dimming due to dust is more or less important than the brightening
due to the increase in the column density of stars induced by inclining the
disk. This ratio can change up to 1.5 magnitudes due to the effect of dust.
The ratio of the apparent to intrinsic axis ratio is not strongly affected by
dust, but has an opposite dependence on inclination according to whether the
lines of sight through the disk are predominantly optically thin or optically
thick. In the former case the vertical distribution of stars makes the galaxy 
appear progressively rounder with increasing inclination than an infinitely 
thin disk, whereas in the latter case the opposite trend is seen, because an 
increasing proportion of the
observed light originates from a thin layer of stars above the dust. 

Assuming that the basic geometry of dust and stars in local universe
  spiral galaxies also applies for higher redshift spiral galaxies, our 
tabulated corrections can be used to correct for the increase in the 
apparent disk scalelength with increasing redshift due to the expected 
increase in opacity.  This will allow the intrinsic evolution of
disk sizes with cosmological epoch to be investigated. As an example, we show 
that for a possible
variation in opacity between 2 at $z=0$ to 8 or more at 
$z=1$, the apparent scalelength in the B band  of a galaxy seen at 
$i=37^{\circ}$ 
 would increase by a factor of at least 1.24 due to dust.

We used our model to analyse the distribution of observed scalelength ratios
between B and I for a sample of disk-dominated spiral galaxies. We found that
the predominance of galaxies with larger apparent scalelength in B than in I is
primarily due to the effects of dust. 
%%%%%%%%%%%%%%%%%%%%%%%%%%%%%%%%%%%%%%%%%%%%%%%%%%%%%%%%%%%%%%%%%%%%%%%
%%%%%%%%%%%%%%%%%%%%%%%%%%%%%%%%%%%%%%%%%%%%%%%%%%%%%%%%%%%%%%%%%%%%%%%
\begin{acknowledgements}
We would like to acknowledge the anonymous referee for his/her perceptive and
constructive criticism which helped us to improve the manuscript.
\end{acknowledgements}

%%%%%%%%%%%%%%%%%%%%%%%%%%%%%%%%%%%%%%%%%%%%%%%%%%%%%%%%%%%%%%%%%%%%%%%
%%%%%%%%%%%%%%%%%%%%%%%%%%%%%%%%%%%%%%%%%%%%%%%%%%%%%%%%%%%%%%%%%%%%%%%

%%%%%%%%%%%%%%%%%%%%%%%%%%%%%%%%%%%%%%%%%%%%%%%%%%%%%%%%%%%%%
%  Hier tab1.tex einfuegen
%%%%%%%%%%%%%%%%%%%%%%%%%%%%%%%%%%%%%%%%%%%%%%%%%%%%%%%%%%
%  Table : tablb2.tex                  16.8.05
%%%%%%%%%%%%%%%%%%%%%%%%%%%%%%%%%%%%%%%%%%%%%%%%%%%%%%%%%
\begin{table}
\caption[]{Structural parameters of the disks in B.}
\label{tablb}
%\vspace{0.5cm}
\[
\begin{array}{|cc|cccr|}
\hline
         &       &         &         &      &  \\
 i &\quad  {\tau}^{f}_{\rm B} \quad 
& \quad Q_{app} & \frac{\displaystyle R_{\rm app}}{\displaystyle R_{0}} & 
\frac{\displaystyle I_{\rm app}^{\rm c}}{\displaystyle I_{\rm 0}^{\rm c}} & \Delta SB \\
 
         &       &         &         &      &  \\
\hline
% MIDAS-Table : paptbl2b.tbl
%Inclin  Tauc     Q_d     Rfit/Rin  Ifit/Iin  DeltaSB  
%------ ------- --------- --------- --------- ---------
     0   & 0.1   & 1.000   & 0.995   & 1.005   &-0.001 \\
     0   & 0.3   & 1.000   & 1.002   & 0.997   & 0.020 \\
     0   & 0.5   & 1.000   & 1.013   & 0.977   & 0.056 \\
     0   & 1.0   & 1.000   & 1.048   & 0.907   & 0.171 \\
     0   & 2.0   & 1.000   & 1.129   & 0.765   & 0.428 \\
     0   & 4.0   & 1.000   & 1.276   & 0.559   & 0.815 \\
     0   & 8.0   & 1.000   & 1.438   & 0.378   & 1.161 \\
          &       &         &         &      &         \\          
    26   & 0.1   & 0.899   & 0.996   & 1.125   &-0.110 \\
    26   & 0.3   & 0.899   & 1.004   & 1.109   &-0.081 \\
    26   & 0.5   & 0.898   & 1.016   & 1.082   &-0.039 \\
    26   & 1.0   & 0.897   & 1.055   & 0.997   & 0.093 \\
    26   & 2.0   & 0.895   & 1.147   & 0.820   & 0.380 \\
    26   & 4.0   & 0.890   & 1.306   & 0.585   & 0.776 \\
    26   & 8.0   & 0.882   & 1.456   & 0.397   & 1.103 \\
          &       &         &         &      &         \\        
    37   & 0.1   & 0.799   & 0.997   & 1.247   &-0.210 \\
    37   & 0.3   & 0.799   & 1.008   & 1.220   &-0.172 \\
    37   & 0.5   & 0.799   & 1.024   & 1.181   &-0.120 \\
    37   & 1.0   & 0.798   & 1.070   & 1.068   & 0.031 \\
    37   & 2.0   & 0.795   & 1.166   & 0.866   & 0.321 \\
    37   & 4.0   & 0.788   & 1.316   & 0.621   & 0.696 \\
    37   & 8.0   & 0.778   & 1.448   & 0.428   & 1.012 \\
          &       &         &         &      &         \\        
    46   & 0.1   & 0.701   & 0.999   & 1.426   &-0.343 \\
    46   & 0.3   & 0.701   & 1.014   & 1.386   &-0.294 \\
    46   & 0.5   & 0.700   & 1.034   & 1.331   &-0.232 \\
    46   & 1.0   & 0.699   & 1.087   & 1.184   &-0.060 \\
    46   & 2.0   & 0.694   & 1.192   & 0.938   & 0.256 \\
    46   & 4.0   & 0.685   & 1.345   & 0.662   & 0.626 \\
    46   & 8.0   & 0.676   & 1.465   & 0.458   & 0.932 \\
          &       &         &         &      &         \\        
    53   & 0.1   & 0.604   & 1.002   & 1.644   &-0.483 \\
    53   & 0.3   & 0.603   & 1.021   & 1.580   &-0.420 \\
    53   & 0.5   & 0.603   & 1.043   & 1.502   &-0.346 \\
    53   & 1.0   & 0.601   & 1.103   & 1.316   &-0.154 \\
    53   & 2.0   & 0.596   & 1.216   & 1.022   & 0.171 \\
    53   & 4.0   & 0.587   & 1.365   & 0.719   & 0.531 \\
    53   & 8.0   & 0.578   & 1.477   & 0.499   & 0.841 \\
          &       &         &         &      &         \\        
    60   & 0.1   & 0.507   & 1.007   & 1.927   &-0.637 \\
    60   & 0.3   & 0.506   & 1.032   & 1.822   &-0.553 \\
    60   & 0.5   & 0.506   & 1.060   & 1.708   &-0.459 \\
    60   & 1.0   & 0.503   & 1.134   & 1.446   &-0.231 \\
    60   & 2.0   & 0.498   & 1.259   & 1.089   & 0.103 \\
    60   & 4.0   & 0.490   & 1.403   & 0.760   & 0.441 \\
    60   & 8.0   & 0.483   & 1.480   & 0.540   & 0.737 \\
          &       &         &         &      &         \\        
    66   & 0.1   & 0.410   & 1.017   & 2.366   &-0.836 \\
    66   & 0.3   & 0.409   & 1.056   & 2.161   &-0.706 \\
    66   & 0.5   & 0.408   & 1.095   & 1.974   &-0.583 \\
    66   & 1.0   & 0.405   & 1.181   & 1.609   &-0.324 \\
    66   & 2.0   & 0.400   & 1.309   & 1.183   & 0.002 \\
    66   & 4.0   & 0.392   & 1.422   & 0.841   & 0.315 \\
    66   & 8.0   & 0.388   & 1.475   & 0.606   & 0.626 \\
          &       &         &         &      &         \\        
    73   & 0.1   & 0.315   & 1.028   & 2.982   &-1.055 \\
    73   & 0.3   & 0.314   & 1.077   & 2.651   &-0.893 \\
    73   & 0.5   & 0.312   & 1.124   & 2.369   &-0.748 \\
    73   & 1.0   & 0.310   & 1.222   & 1.866   &-0.469 \\
    73   & 2.0   & 0.304   & 1.349   & 1.350   &-0.140 \\
    73   & 4.0   & 0.298   & 1.453   & 0.956   & 0.177 \\
    73   & 8.0   & 0.295   & 1.492   & 0.687   & 0.499 \\
%----------------------------------------------------------
\hline
\end{array}
\]
\end{table}
%=========================================================================

%%%%%%%%%%%%%%%%%%%%%%%%%%%%%%%%%%%%%%%%%%%%%%%%%%%%%%%%%%
%  Table : tablv2.tex                  16.8.05
%%%%%%%%%%%%%%%%%%%%%%%%%%%%%%%%%%%%%%%%%%%%%%%%%%%%%%%%%
\begin{table}
\caption[]{Structural parameters of the disks in V.}
\label{tablb}
%\vspace{0.5cm}
\[
\begin{array}{|cc|cccr|}
\hline
         &       &         &         &      &  \\
 i &\quad {\tau}^{f}_{\rm B} \quad 
& \quad Q_{app} & \frac{\displaystyle R_{\rm app}}{\displaystyle R_{0}} & 
\frac{\displaystyle I_{\rm app}^{\rm c}}{\displaystyle I_{\rm 0}^{\rm c}} &
         \Delta SB \\
 
         &       &         &         &      &  \\
\hline
% MIDAS-Table : paptbl2v.tbl
%Inclin  Tauc     Q_d     Rfit/Rin  Ifit/Iin  DeltaSB  
%------------------------------------------------------
    0   & 0.1   & 1.000   & 0.996   & 1.004   &-0.001 \\ 
    0   & 0.3   & 1.000   & 0.999   & 1.002   & 0.010 \\ 
    0   & 0.5   & 1.000   & 1.006   & 0.991   & 0.032 \\ 
    0   & 1.0   & 1.000   & 1.029   & 0.947   & 0.108 \\ 
    0   & 2.0   & 1.000   & 1.087   & 0.837   & 0.296 \\ 
    0   & 4.0   & 1.000   & 1.208   & 0.647   & 0.644 \\ 
    0   & 8.0   & 1.000   & 1.376   & 0.443   & 1.031 \\ 
         &       &         &         &      &         \\       
   26   & 0.1   & 0.900   & 0.996   & 1.125   &-0.112 \\ 
   26   & 0.3   & 0.900   & 1.001   & 1.118   &-0.095 \\ 
   26   & 0.5   & 0.900   & 1.010   & 1.101   &-0.067 \\ 
   26   & 1.0   & 0.899   & 1.037   & 1.041   & 0.024 \\ 
   26   & 2.0   & 0.898   & 1.104   & 0.902   & 0.242 \\ 
   26   & 4.0   & 0.894   & 1.235   & 0.678   & 0.612 \\ 
   26   & 8.0   & 0.888   & 1.401   & 0.461   & 0.974 \\ 
         &       &         &         &      &         \\     
   37   & 0.1   & 0.800   & 0.998   & 1.248   &-0.212 \\ 
   37   & 0.3   & 0.800   & 1.005   & 1.233   &-0.189 \\ 
   37   & 0.5   & 0.800   & 1.016   & 1.207   &-0.155 \\ 
   37   & 1.0   & 0.799   & 1.048   & 1.127   &-0.050 \\ 
   37   & 2.0   & 0.797   & 1.118   & 0.964   & 0.174 \\ 
   37   & 4.0   & 0.793   & 1.246   & 0.724   & 0.530 \\ 
   37   & 8.0   & 0.785   & 1.395   & 0.499   & 0.883 \\ 
         &       &         &         &      &         \\     
   46   & 0.1   & 0.702   & 1.000   & 1.426   &-0.344 \\   
   46   & 0.3   & 0.702   & 1.010   & 1.400   &-0.312 \\ 
   46   & 0.5   & 0.702   & 1.023   & 1.362   &-0.268 \\ 
   46   & 1.0   & 0.701   & 1.062   & 1.251   &-0.143 \\ 
   46   & 2.0   & 0.698   & 1.144   & 1.043   & 0.111 \\ 
   46   & 4.0   & 0.692   & 1.282   & 0.765   & 0.474 \\ 
   46   & 8.0   & 0.683   & 1.418   & 0.529   & 0.805 \\ 
         &       &         &         &      &         \\     
   53   & 0.1   & 0.604   & 1.002   & 1.644   &-0.484 \\ 
   53   & 0.3   & 0.604   & 1.015   & 1.601   &-0.441 \\ 
   53   & 0.5   & 0.604   & 1.031   & 1.545   &-0.388 \\ 
   53   & 1.0   & 0.603   & 1.075   & 1.400   &-0.245 \\ 
   53   & 2.0   & 0.600   & 1.164   & 1.143   & 0.028 \\ 
   53   & 4.0   & 0.594   & 1.306   & 0.826   & 0.389 \\ 
   53   & 8.0   & 0.586   & 1.434   & 0.572   & 0.715 \\ 
         &       &         &         &      &         \\     
   60   & 0.1   & 0.508   & 1.007   & 1.933   &-0.642 \\ 
   60   & 0.3   & 0.508   & 1.024   & 1.857   &-0.581 \\ 
   60   & 0.5   & 0.507   & 1.046   & 1.770   &-0.512 \\ 
   60   & 1.0   & 0.506   & 1.101   & 1.557   &-0.336 \\ 
   60   & 2.0   & 0.503   & 1.202   & 1.231   &-0.042 \\ 
   60   & 4.0   & 0.496   & 1.339   & 0.880   & 0.304 \\ 
   60   & 8.0   & 0.489   & 1.444   & 0.618   & 0.609 \\ 
         &       &         &         &      &         \\     
   66   & 0.1   & 0.412   & 1.015   & 2.379   &-0.844 \\ 
   66   & 0.3   & 0.411   & 1.044   & 2.224   &-0.747 \\ 
   66   & 0.5   & 0.410   & 1.074   & 2.073   &-0.651 \\ 
   66   & 1.0   & 0.408   & 1.143   & 1.757   &-0.437 \\ 
   66   & 2.0   & 0.404   & 1.253   & 1.343   &-0.131 \\ 
   66   & 4.0   & 0.398   & 1.378   & 0.958   & 0.193 \\ 
   66   & 8.0   & 0.393   & 1.451   & 0.687   & 0.494 \\ 
         &       &         &         &      &         \\     
   73   & 0.1   & 0.316   & 1.025   & 3.006   &-1.066 \\ 
   73   & 0.3   & 0.315   & 1.061   & 2.747   &-0.943 \\ 
   73   & 0.5   & 0.315   & 1.097   & 2.514   &-0.827 \\ 
   73   & 1.0   & 0.312   & 1.178   & 2.062   &-0.586 \\ 
   73   & 2.0   & 0.308   & 1.295   & 1.536   &-0.272 \\ 
   73   & 4.0   & 0.302   & 1.410   & 1.089   & 0.053 \\ 
   73   & 8.0   & 0.298   & 1.470   & 0.781   & 0.364 \\ 
%----------------------------------------------------------
\hline
\end{array}
\]
\end{table}
%=========================================================================

%%%%%%%%%%%%%%%%%%%%%%%%%%%%%%%%%%%%%%%%%%%%%%%%%%%%%%%%%%
%  Table : tabli2.tex                  16.8.05
%%%%%%%%%%%%%%%%%%%%%%%%%%%%%%%%%%%%%%%%%%%%%%%%%%%%%%%%%
\begin{table}
\caption[]{Structural parameters of the disks in I.}
\label{tabli}
%\vspace{0.5cm}
\[
\begin{array}{|cc|cccr|}
\hline
         &       &         &         &      &  \\
 i &\quad {\tau}^{f}_{\rm B} \quad 
& \quad Q_{app} & \frac{\displaystyle R_{\rm app}}{\displaystyle R_{0}} & 
\frac{\displaystyle I_{\rm app}^{\rm c}}{\displaystyle I_{\rm 0}^{\rm c}} & \Delta SB \\
 
         &       &         &         &      &  \\
\hline
% MIDAS-Table : paptbl2i.tbl
%Inclin  Tauc     Q_d     Rfit/Rin  Ifit/Iin  DeltaSB  
%------------------------------------------------------
     0   & 0.1   & 1.000   & 0.996   & 1.007   &-0.006 \\ 
     0   & 0.3   & 1.000   & 0.998   & 1.007   &-0.001 \\ 
     0   & 0.5   & 1.000   & 1.000   & 1.003   & 0.009 \\ 
     0   & 1.0   & 1.000   & 1.011   & 0.982   & 0.046 \\ 
     0   & 2.0   & 1.000   & 1.043   & 0.919   & 0.148 \\ 
     0   & 4.0   & 1.000   & 1.117   & 0.780   & 0.379 \\ 
     0   & 8.0   & 1.000   & 1.248   & 0.575   & 0.751 \\ 
         &       &         &         &      &          \\      
    26   & 0.1   & 0.900   & 1.003   & 1.116   &-0.103 \\ 
    26   & 0.3   & 0.900   & 1.006   & 1.114   &-0.095 \\ 
    26   & 0.5   & 0.900   & 1.011   & 1.107   &-0.082 \\ 
    26   & 1.0   & 0.899   & 1.026   & 1.075   &-0.034 \\ 
    26   & 2.0   & 0.899   & 1.065   & 0.989   & 0.089 \\ 
    26   & 4.0   & 0.899   & 1.147   & 0.818   & 0.345 \\ 
    26   & 8.0   & 0.897   & 1.282   & 0.590   & 0.717 \\ 
         &       &         &         &      &          \\      
    37   & 0.1   & 0.803   & 0.998   & 1.253   &-0.218 \\ 
    37   & 0.3   & 0.803   & 1.002   & 1.246   &-0.206 \\ 
    37   & 0.5   & 0.803   & 1.008   & 1.233   &-0.187 \\ 
    37   & 1.0   & 0.803   & 1.026   & 1.186   &-0.128 \\ 
    37   & 2.0   & 0.802   & 1.068   & 1.077   & 0.011 \\ 
    37   & 4.0   & 0.801   & 1.156   & 0.878   & 0.286 \\ 
    37   & 8.0   & 0.798   & 1.293   & 0.630   & 0.655 \\ 
         &       &         &         &      &          \\      
    46   & 0.1   & 0.705   & 1.003   & 1.419   &-0.338 \\ 
    46   & 0.3   & 0.705   & 1.008   & 1.404   &-0.319 \\ 
    46   & 0.5   & 0.705   & 1.015   & 1.384   &-0.297 \\ 
    46   & 1.0   & 0.705   & 1.036   & 1.320   &-0.227 \\ 
    46   & 2.0   & 0.704   & 1.086   & 1.179   &-0.069 \\ 
    46   & 4.0   & 0.701   & 1.184   & 0.938   & 0.221 \\ 
    46   & 8.0   & 0.697   & 1.321   & 0.664   & 0.584 \\ 
         &       &         &         &      &          \\      
    53   & 0.1   & 0.608   & 1.005   & 1.641   &-0.478 \\ 
    53   & 0.3   & 0.608   & 1.013   & 1.614   &-0.453 \\ 
    53   & 0.5   & 0.608   & 1.022   & 1.580   &-0.422 \\ 
    53   & 1.0   & 0.607   & 1.049   & 1.485   &-0.336 \\ 
    53   & 2.0   & 0.606   & 1.106   & 1.296   &-0.155 \\ 
    53   & 4.0   & 0.604   & 1.211   & 1.008   & 0.157 \\ 
    53   & 8.0   & 0.599   & 1.344   & 0.709   & 0.513 \\ 
         &       &         &         &      &          \\      
    60   & 0.1   & 0.512   & 1.007   & 1.939   &-0.641 \\ 
    60   & 0.3   & 0.512   & 1.019   & 1.890   &-0.604 \\ 
    60   & 0.5   & 0.512   & 1.032   & 1.834   &-0.563 \\ 
    60   & 1.0   & 0.511   & 1.066   & 1.690   &-0.453 \\ 
    60   & 2.0   & 0.509   & 1.132   & 1.437   &-0.246 \\ 
    60   & 4.0   & 0.506   & 1.240   & 1.094   & 0.068 \\ 
    60   & 8.0   & 0.501   & 1.363   & 0.770   & 0.405 \\ 
         &       &         &         &      &          \\      
    66   & 0.1   & 0.415   & 1.018   & 2.362   &-0.832 \\ 
    66   & 0.3   & 0.415   & 1.034   & 2.268   &-0.778 \\ 
    66   & 0.5   & 0.415   & 1.051   & 2.173   &-0.722 \\ 
    66   & 1.0   & 0.414   & 1.093   & 1.950   &-0.584 \\ 
    66   & 2.0   & 0.412   & 1.171   & 1.598   &-0.348 \\ 
    66   & 4.0   & 0.408   & 1.284   & 1.179   &-0.024 \\ 
    66   & 8.0   & 0.403   & 1.387   & 0.833   & 0.301 \\ 
         &       &         &         &      &          \\      
    73   & 0.1   & 0.320   & 1.024   & 3.012   &-1.066 \\ 
    73   & 0.3   & 0.320   & 1.048   & 2.831   &-0.985 \\ 
    73   & 0.5   & 0.320   & 1.072   & 2.663   &-0.905 \\ 
    73   & 1.0   & 0.318   & 1.128   & 2.308   &-0.729 \\ 
    73   & 2.0   & 0.316   & 1.218   & 1.822   &-0.461 \\ 
    73   & 4.0   & 0.312   & 1.329   & 1.326   &-0.142 \\ 
    73   & 8.0   & 0.307   & 1.411   & 0.948   & 0.167 \\ 
%----------------------------------------------------------
\hline
\end{array}
\]
\end{table}
%=========================================================================

%%%%%%%%%%%%%%%%%%%%%%%%%%%%%%%%%%%%%%%%%%%%%%%%%%%%%%%%%%
%  Table : tablj2.tex                  16.8.05
%%%%%%%%%%%%%%%%%%%%%%%%%%%%%%%%%%%%%%%%%%%%%%%%%%%%%%%%%
\begin{table}
\caption[]{Structural parameters of the disks in J.}
\label{tablj}
%\vspace{0.5cm}
\[
\begin{array}{|cc|cccr|}
\hline
         &       &         &         &      &  \\
 i &\quad {\tau}^{f}_{\rm B} \quad 
& \quad Q_{app} & \frac{\displaystyle R_{\rm app}}{\displaystyle R_{0}} & 
\frac{\displaystyle I_{\rm app}^{\rm c}}{\displaystyle I_{\rm 0}^{\rm c}} & \Delta SB \\
 
         &       &         &         &      &  \\
\hline
% MIDAS-Table : paptbl2j.tbl
%Inclin  Tauc     Q_d     Rfit/Rin  Ifit/Iin  DeltaSB  
%------------------------------------------------------
     0   & 0.1   & 1.000   & 1.001   & 1.000   & 0.001   \\ 
     0   & 0.3   & 1.000   & 1.003   & 0.999   & 0.005   \\ 
     0   & 0.5   & 1.000   & 1.005   & 0.996   & 0.011   \\ 
     0   & 1.0   & 1.000   & 1.012   & 0.985   & 0.030   \\ 
     0   & 2.0   & 1.000   & 1.027   & 0.952   & 0.079   \\ 
     0   & 4.0   & 1.000   & 1.061   & 0.877   & 0.193   \\ 
     0   & 8.0   & 1.000   & 1.131   & 0.737   & 0.420   \\ 
         &       &         &         &      &           \\     
    26   & 0.1   & 0.902   & 1.002   & 1.119   &-0.106  \\ 
    26   & 0.3   & 0.902   & 1.004   & 1.115   &-0.099  \\ 
    26   & 0.5   & 0.902   & 1.007   & 1.110   &-0.091  \\ 
    26   & 1.0   & 0.902   & 1.015   & 1.092   &-0.065  \\ 
    26   & 2.0   & 0.901   & 1.033   & 1.047   &-0.005  \\ 
    26   & 4.0   & 0.901   & 1.075   & 0.947   & 0.132   \\ 
    26   & 8.0   & 0.900   & 1.158   & 0.772   & 0.392   \\ 
         &      &        &        &     &           \\     
    37   & 0.1   & 0.804   & 1.003   & 1.239   &-0.202  \\ 
    37   & 0.3   & 0.804   & 1.006   & 1.233   &-0.194  \\ 
    37   & 0.5   & 0.804   & 1.009   & 1.226   &-0.185  \\ 
    37   & 1.0   & 0.804   & 1.018   & 1.203   &-0.157  \\ 
    37   & 2.0   & 0.804   & 1.038   & 1.147   &-0.091  \\ 
    37   & 4.0   & 0.803   & 1.081   & 1.030   & 0.050   \\ 
    37   & 8.0   & 0.803   & 1.161   & 0.837   & 0.308   \\ 
         &      &        &        &     &           \\     
    46   & 0.1   & 0.707   & 1.005   & 1.415   &-0.330  \\ 
    46   & 0.3   & 0.707   & 1.009   & 1.403   &-0.319  \\ 
    46   & 0.5   & 0.708   & 1.013   & 1.391   &-0.306  \\ 
    46   & 1.0   & 0.707   & 1.024   & 1.354   &-0.268  \\ 
    46   & 2.0   & 0.707   & 1.049   & 1.274   &-0.186  \\ 
    46   & 4.0   & 0.706   & 1.099   & 1.121   &-0.021  \\ 
    46   & 8.0   & 0.704   & 1.189   & 0.886   & 0.261   \\ 
         &      &        &        &     &           \\     
    53   & 0.1   & 0.610   & 1.007   & 1.633   &-0.470  \\ 
    53   & 0.3   & 0.610   & 1.012   & 1.613   &-0.453  \\ 
    53   & 0.5   & 0.610   & 1.017   & 1.593   &-0.436  \\ 
    53   & 1.0   & 0.610   & 1.030   & 1.540   &-0.390  \\ 
    53   & 2.0   & 0.610   & 1.058   & 1.432   &-0.294  \\ 
    53   & 4.0   & 0.609   & 1.115   & 1.237   &-0.109  \\ 
    53   & 8.0   & 0.607   & 1.211   & 0.957   & 0.188   \\ 
         &      &        &        &     &           \\     
    60   & 0.1   & 0.515   & 1.011   & 1.920   &-0.627  \\ 
    60   & 0.3   & 0.514   & 1.017   & 1.891   &-0.606  \\ 
    60   & 0.5   & 0.514   & 1.024   & 1.861   &-0.585  \\ 
    60   & 1.0   & 0.514   & 1.040   & 1.782   &-0.528  \\ 
    60   & 2.0   & 0.514   & 1.074   & 1.630   &-0.414  \\ 
    60   & 4.0   & 0.512   & 1.137   & 1.376   &-0.208  \\ 
    60   & 8.0   & 0.510   & 1.237   & 1.044   & 0.097   \\ 
         &      &        &        &     &           \\     
    66   & 0.1   & 0.419   & 1.017   & 2.365   &-0.832  \\ 
    66   & 0.3   & 0.419   & 1.026   & 2.305   &-0.797  \\ 
    66   & 0.5   & 0.419   & 1.036   & 2.245   &-0.762  \\ 
    66   & 1.0   & 0.418   & 1.060   & 2.103   &-0.677  \\ 
    66   & 2.0   & 0.418   & 1.107   & 1.855   &-0.522  \\ 
    66   & 4.0   & 0.415   & 1.185   & 1.495   &-0.277  \\ 
    66   & 8.0   & 0.412   & 1.286   & 1.102   & 0.031   \\ 
         &      &        &        &     &           \\     
    73   & 0.1   & 0.325   & 1.025   & 2.992   &-1.056  \\ 
    73   & 0.3   & 0.325   & 1.037   & 2.894   &-1.013  \\ 
    73   & 0.5   & 0.325   & 1.048   & 2.800   &-0.970  \\ 
    73   & 1.0   & 0.324   & 1.077   & 2.584   &-0.870  \\ 
    73   & 2.0   & 0.323   & 1.130   & 2.230   &-0.694  \\ 
    73   & 4.0   & 0.321   & 1.215   & 1.752   &-0.433  \\ 
    73   & 8.0   & 0.316   & 1.315   & 1.270   &-0.116  \\ 
%----------------------------------------------------------
\hline
\end{array}
\]
\end{table}
%=========================================================================

%%%%%%%%%%%%%%%%%%%%%%%%%%%%%%%%%%%%%%%%%%%%%%%%%%%%%%%%%%
%  Table : tablk2.tex                  16.8.05
%%%%%%%%%%%%%%%%%%%%%%%%%%%%%%%%%%%%%%%%%%%%%%%%%%%%%%%%%
\begin{table}
\caption[]{Structural parameters of the disks in K.}
\label{tablk}
%\vspace{0.5cm}
\[
\begin{array}{|cc|cccr|}
\hline
         &       &         &         &      &  \\
 i &\quad {\tau}^{f}_{\rm B} \quad 
& \quad Q_{app} & \frac{\displaystyle R_{\rm app}}{\displaystyle R_{0}} & 
\frac{\displaystyle I_{\rm app}^{\rm c}}{\displaystyle I_{\rm 0}^{\rm c}} & \Delta SB \\
 
         &       &         &         &      &  \\
\hline
% MIDAS-Table : paptbl2k.tbl
%Inclin  Tauc     Q_d     Rfit/Rin  Ifit/Iin  DeltaSB  
%------------------------------------------------------
     0   & 0.1   & 1.000   & 1.002   & 0.998   & 0.002 \\ 
     0   & 0.3   & 1.000   & 1.003   & 0.995   & 0.006 \\ 
     0   & 0.5   & 1.000   & 1.005   & 0.992   & 0.011 \\ 
     0   & 1.0   & 1.000   & 1.008   & 0.984   & 0.022 \\ 
     0   & 2.0   & 1.000   & 1.014   & 0.967   & 0.045 \\ 
     0   & 4.0   & 1.000   & 1.027   & 0.933   & 0.091 \\ 
     0   & 8.0   & 1.000   & 1.054   & 0.865   & 0.188 \\ 
         &       &         &         &      &          \\        
    26   & 0.1   & 0.903   & 1.003   & 1.117   &-0.101 \\ 
    26   & 0.3   & 0.903   & 1.005   & 1.112   &-0.096 \\ 
    26   & 0.5   & 0.903   & 1.006   & 1.108   &-0.090 \\ 
    26   & 1.0   & 0.903   & 1.010   & 1.096   &-0.076 \\ 
    26   & 2.0   & 0.903   & 1.018   & 1.072   &-0.047 \\ 
    26   & 4.0   & 0.902   & 1.035   & 1.024   & 0.012 \\ 
    26   & 8.0   & 0.902   & 1.067   & 0.933   & 0.127 \\ 
         &       &         &         &      &          \\        
    37   & 0.1   & 0.806   & 1.004   & 1.236   &-0.195 \\ 
    37   & 0.3   & 0.806   & 1.006   & 1.231   &-0.189 \\ 
    37   & 0.5   & 0.806   & 1.007   & 1.225   &-0.184 \\ 
    37   & 1.0   & 0.806   & 1.011   & 1.212   &-0.170 \\ 
    37   & 2.0   & 0.806   & 1.019   & 1.186   &-0.142 \\ 
    37   & 4.0   & 0.806   & 1.034   & 1.134   &-0.085 \\ 
    37   & 8.0   & 0.806   & 1.066   & 1.036   & 0.028 \\ 
         &       &         &         &      &          \\        
    46   & 0.1   & 0.710   & 1.006   & 1.410   &-0.321 \\ 
    46   & 0.3   & 0.710   & 1.008   & 1.402   &-0.314 \\ 
    46   & 0.5   & 0.710   & 1.010   & 1.395   &-0.307 \\ 
    46   & 1.0   & 0.709   & 1.015   & 1.375   &-0.289 \\ 
    46   & 2.0   & 0.709   & 1.025   & 1.337   &-0.253 \\ 
    46   & 4.0   & 0.709   & 1.045   & 1.263   &-0.182 \\ 
    46   & 8.0   & 0.709   & 1.083   & 1.129   &-0.045 \\ 
         &       &         &         &      &          \\        
    53   & 0.1   & 0.613   & 1.008   & 1.627   &-0.459 \\ 
    53   & 0.3   & 0.613   & 1.010   & 1.617   &-0.450 \\ 
    53   & 0.5   & 0.613   & 1.013   & 1.606   &-0.442 \\ 
    53   & 1.0   & 0.613   & 1.019   & 1.580   &-0.421 \\ 
    53   & 2.0   & 0.613   & 1.030   & 1.528   &-0.379 \\ 
    53   & 4.0   & 0.613   & 1.053   & 1.430   &-0.297 \\ 
    53   & 8.0   & 0.613   & 1.096   & 1.260   &-0.144 \\ 
         &       &         &         &      &          \\        
    60   & 0.1   & 0.519   & 1.012   & 1.910   &-0.612 \\ 
    60   & 0.3   & 0.518   & 1.014   & 1.895   &-0.602 \\ 
    60   & 0.5   & 0.518   & 1.017   & 1.881   &-0.593 \\ 
    60   & 1.0   & 0.518   & 1.023   & 1.845   &-0.569 \\ 
    60   & 2.0   & 0.518   & 1.036   & 1.776   &-0.521 \\ 
    60   & 4.0   & 0.518   & 1.061   & 1.647   &-0.429 \\ 
    60   & 8.0   & 0.517   & 1.109   & 1.431   &-0.261 \\ 
         &       &         &         &      &          \\        
    66   & 0.1   & 0.424   & 1.017   & 2.350   &-0.817 \\ 
    66   & 0.3   & 0.424   & 1.021   & 2.323   &-0.802 \\ 
    66   & 0.5   & 0.424   & 1.025   & 2.296   &-0.787 \\ 
    66   & 1.0   & 0.424   & 1.035   & 2.231   &-0.750 \\ 
    66   & 2.0   & 0.424   & 1.053   & 2.110   &-0.679 \\ 
    66   & 4.0   & 0.423   & 1.089   & 1.898   &-0.549 \\ 
    66   & 8.0   & 0.422   & 1.150   & 1.578   &-0.337 \\ 
         &       &         &         &      &          \\        
    73   & 0.1   & 0.331   & 1.026   & 2.960   &-1.036 \\ 
    73   & 0.3   & 0.331   & 1.031   & 2.919   &-1.019 \\ 
    73   & 0.5   & 0.331   & 1.035   & 2.881   &-1.002 \\ 
    73   & 1.0   & 0.331   & 1.046   & 2.787   &-0.961 \\ 
    73   & 2.0   & 0.330   & 1.067   & 2.614   &-0.882 \\ 
    73   & 4.0   & 0.330   & 1.106   & 2.320   &-0.739 \\ 
    73   & 8.0   & 0.328   & 1.173   & 1.892   &-0.512 \\ 
%----------------------------------------------------------
\hline
\end{array}
\]
\end{table}
%=========================================================================

%%%%%%%%%%%%%%%%%%%%%%%%%%%%%%%%%%%%%%%%%%%%%%%%%%%%%%%%%%%%%

\end{document}